\documentclass[aps,floats,amssymb,amsmath,prb,nofootinbib,twocolumn,superscriptaddress]{revtex4-1}

\usepackage{booktabs}
\usepackage{calc}
\usepackage{psfrag}
\usepackage{graphicx}
\usepackage{color}
\usepackage[dvipsnames]{xcolor}
\usepackage[unicode=true,pdfusetitle,
 bookmarks=true,bookmarksnumbered=false,bookmarksopen=false,
 breaklinks=false,pdfborder={0 0 0},backref=false,colorlinks=true]
 {hyperref}
\usepackage{geometry}
\pdfpageattr {/Group << /S /Transparency /I true /CS /DeviceRGB>>} %necessary for transparent plots

\newcommand{\blue}{\color{blue}}
\newcommand{\red}{\color{red}}
\newcommand{\up}{\uparrow}
\newcommand{\dn}{\downarrow}

\usepackage{pdfpages}

\makeatletter
\AtBeginDocument{\let\LS@rot\@undefined}
\makeatother

\begin{document}

\title{Analytical solution for time-integrals in diagrammatic expansions: application to real-frequency diagrammatic Monte Carlo}
\author{J. Vu\v ci\v cevi\'c}
\affiliation{Scientific Computing Laboratory, Center for the Study of Complex Systems, Institute of Physics Belgrade,
University of Belgrade, Pregrevica 118, 11080 Belgrade, Serbia}
\author{P. Stipsi\'c}
\affiliation{Scientific Computing Laboratory, Center for the Study of Complex Systems, Institute of Physics Belgrade,
University of Belgrade, Pregrevica 118, 11080 Belgrade, Serbia}
\affiliation{Faculty of Physics, University of Belgrade, Studentski trg 12, 11001 Belgrade, Serbia}
\author{M. Ferrero}
\affiliation{CPHT, CNRS, Ecole Polytechnique, Institut Polytechnique de Paris, Route de Saclay, 91128 Palaiseau, France}
\affiliation{Coll\`ege de France, 11 place Marcelin Berthelot, 75005 Paris, France}

\begin{abstract}

The past years have seen a revived interest in the diagrammatic Monte Carlo
(DiagMC) methods for interacting fermions on a lattice.  A promising recent
development allows one to now circumvent the analytical continuation of dynamic
observables in DiagMC calculations within the Matsubara formalism.  This is
made possible by symbolic algebra algorithms, which can be used to analytically
solve the internal Matsubara frequency summations of Feynman diagrams.  In
this paper, we take a different approach and show that it yields improved results. We present a
closed-form analytical solution of imaginary-time integrals that appear in the
time-domain formulation of Feynman diagrams.  We implement and test a DiagMC
algorithm based on this analytical solution and show that it has numerous
significant advantages.
Most importantly, the algorithm is general enough for any kind of single-time
correlation function series, involving any single-particle vertex insertions.
Therefore, it readily allows for the use of action-shifted schemes, aimed at
improving the convergence properties of the series.  By performing a
frequency-resolved action-shift tuning, we are able to further improve the
method and converge the self-energy in a non-trivial regime, with only 3-4
perturbation orders.
Finally, we identify time integrals of the same general form in many commonly
used Monte Carlo algorithms and therefore expect a broader usage of our
analytical solution.

\end{abstract}

\maketitle

Finding controlled solutions of the Hubbard model is one of the central
challenges in condensed matter
physics\cite{NatPhysEditorial2013,leblancPRX2015,RohringerRMP2018,schafer2020tracking}.
Many common approaches to this problem rely on the
stochastic (Monte Carlo) summation of various expansions and decompositions of
relevant physical quantities. However, Monte Carlo (MC) algorithms are often
plagued by two notorious problems: the fermionic sign problem and the
analytical continuation of frequency-dependent quantities in calculations based
on the Matsubara formalism\cite{
SandvikPRB1998,
SyljuaasenPRB2008,
FuchsPRE2010,
GhanemPRB2020} (alternatively the dynamical sign problem in the
Kadanoff-Baym and Keldysh formalism calculations\cite{
AokiRMP2014,
FreericksPRB2008,
EcksteinPRL2008,
SchiroPRB2009,
WernerPRB2009,
EcksteinPRL2009,
SchiroPRL2010,
WernerPRB2010,
EcksteinPRB2010,
SchiroPRB2011,
WernerPRB2013,
Eckstein2013,
cohen2014a,
cohen2014b,
cohenPRL2015}). In
DiagMC methods\cite{
ProkofevPRL1998,
MishchenkoPRB2000,
Prokofev2007,
ProkofevPRB2008,
GullPRB2010,
KozikEPL2010,
VanhouckePhProc2010,
PolletRPP2012,
VanHouckeNatPhys2012,
KulaginPRL2013,
KulaginPRB2013,
RossiPRL2018,
riccardo_2017,
VanHouckePRB2019,
ChenNatComm2019} (as opposed to determinantal methods such as CTINT or
CTAUX\cite{rubtsov2004,rubtsov_prb_2005,GullEPL2008,gullRMP2011}) an additional
problem is often the slow (or absence of) convergence of the series with
respect to the perturbation order. In recent years, several works have started
to address the problems of obtaining real-frequency
quantities~\cite{
profumoPRB2015,
MoutenetPRB2019,
bertrand2019a,
bertrand2019b,
taheridehkordiPRB2019,
VucicevicPRB2020,
taheridehkordiPRB2020,
taheridehkordiPRB2020a,
MacekPRL2020} and series convergence in
DiagMC~\cite{wuPRB2017,rossiEPL2020,simkovicRPA2020, simkovicPRL2020,KimPRL2020,LenihanArxiv2020}.

In Refs~\onlinecite{profumoPRB2015,wuPRB2017} it has been shown that a
convenient transformation of the interaction-expansion series can be used to
significantly improve its convergence and sometimes allows one to converge the
electronic self-energy with only a few perturbation orders where it would have
otherwise been impossible. The method relies on a transformation of the action
which affects the bare propagator at the cost of an additional expansion, i.e.
more diagram topologies need to be taken into account. Alternatively, this
transformation can be viewed as a Maclaurin expansion of the bare propagator
with respect to a small chemical potential shift. The resulting convergence
speed up comes from an increased convergence radius of the transformed series.

In a separate line of work, DiagMC methods have been proposed that are based on the Matsubara
formalism that do not require an ill-defined analytical
continuation\cite{taheridehkordiPRB2019}. Such methods have so far been implemented for the calculation
of the self-energy\cite{VucicevicPRB2020,taheridehkordiPRB2020} and the dynamical
spin susceptibility\cite{taheridehkordiPRB2020a}. The algorithms differ in
some aspects but all rely on the symbolic algebra solution of the internal
Matsubara frequency summations appearing in Feynman diagrams. However, this approach
has some downsides. First, numerical regulators are needed to properly evaluate
Bose-Einstein distribution functions and diverging ratios that appear in the
analytical expressions, and also poles on the real-axis (effective broadening
of the real-frequency results). In the case of finite cyclic lattice
calculations, multiple precision algebra is needed in order to cancel
divergences even with relatively large regulators.\cite{VucicevicPRB2020} Most
importantly, in the Matsubara summation algorithm, applying the series
transformation from Refs~\onlinecite{profumoPRB2015,wuPRB2017} would require a
separate analytical solution for each of the additional diagram topologies,
that are very numerous, and the calculation would become rather impractical.
More generally, treating any distinct diagram requires that the Matsubara
frequency summations be performed algorithmically beforehand. This makes it
difficult to devise MC sampling algorithms that go to indefinite
perturbation orders,
unless the Matsubara summation part is sufficiently optimized so that it no longer presents
a prohibitive performance penalty if performed at the time of the Monte Carlo sampling.

%because the upper limit of the perturbation orders must be set in advance.

In this paper, we show that it can be advantageous to start from the imaginary-time domain formulation of Feynman diagrams. A diagram contribution then features a multiple imaginary-time integral, rather than sums over Matsubara frequencies. The multiple integral can be solved analytically and we present a general solution. This analytical solution, although equivalent to the analytical Matsubara summation, has a simpler and a more convenient form that does not feature Bose-Einstein distribution functions or diverging ratios. As a result, numerical regulators are not needed and the need for multiple precision arithmetic may arise only at very high perturbation orders. The numerical evaluation yields a sum of poles of various orders on a uniform grid on the real axis. The ability to separate contributions of poles of different orders allows one to formally extract the real-frequency result without any numerical broadening. Finally, the analytical solution is general and applies to all diagram topologies that would appear in the transformed series proposed in Refs~\onlinecite{profumoPRB2015,wuPRB2017} or any other diagrammatic series for single-time correlation functions. This paves the way for real-frequency diagrammatic algorithms formulated in real-space that are not a priori limited to small perturbation orders (similarly to CTINT or CTAUX\cite{gullRMP2011}).

In this work, we apply the analytical time-integral to the momentum-space DiagMC for the calculation of the self-energy, and implement and thoroughly test the method. We reproduce the self-energy results from Ref.\onlinecite{wuPRB2017} and supplement them with real-axis results, free of the uncontrolled systematic error that would otherwise come from the analytical continuation. Furthermore, we show that even if a full convergence is not possible with a single choice of the action-tuning parameter, one can choose the optimal tuning parameter for each frequency independently~\cite{bertrand2019b}. Such a frequency-resolved resummation can be used to improve the solution and in some cases systematically eliminate the non-physical features that appear in the result due to the truncation of the series at a finite order.

The paper is organized as follows. In Section~\ref{sec:model}, we define the
model and the basic assumptions of our calculations. In
Section~\ref{sec:method}, we introduce our method in details. First, in
Section~\ref{sec:analytical_solution}, we present the analytical solution of
the general multiple-time integral that appears in time-domain formulation of
Feynman diagrams and discuss the numerical evaluation of the final expression.
Then in Section~\ref{sec:g0_expansion}, we show the analytical solution for the
Fourier transform of the Maclaurin expansion of the bare propagator, which is
essential for our DiagMC algorithm. In Section~\ref{sec:application_to_diagmc},
we discuss in detail how our analytical solutions can be applied in the context
of DiagMC for the self-energy. In Section~\ref{sec:results}, we discuss our
results and benchmarks and then give closing remarks in
Section~\ref{sec:conclusion}. Additional details of analytical derivations and
further benchmarks and examples of calculations can be found in the appendices.

\section{Model} \label{sec:model}

We solve the Hubbard model given by the Hamiltonian
\begin{equation}
 H = -\sum_{\sigma,ij} t_{ij} c^\dagger_{\sigma,i}c_{\sigma,j}+ U\sum_i n_\up n_\dn - \mu \sum_{\sigma,i} n_{\sigma,i}
\end{equation}
where $\sigma\in\{\up,\dn\}$, $i,j$ enumerate lattice sites, $t_{ij}$ is the hopping amplitude between the sites $i$ and $j$, $U$ is the onsite coupling constant, and $\mu$ is the chemical potential. We only consider the Hubbard model on the square lattice with the nearest-neighbor hopping $t$ and next-nearest-neighbor hopping $t'$. The bare dispersion is given by
\begin{equation}
 \varepsilon_\mathbf{k} = -2t(\cos k_x + \cos k_y)-4t'\cos k_x \cos k_y
\end{equation}
We define $D=4t$, which will be used as the unit of energy unless stated otherwise. We restrict to thermal equilibrium and paramagnetic phases with full lattice symmetry.

\section{Methods} \label{sec:method}

The idea of DiagMC algorithms is to stochastically compute the coefficients of
a perturbation series describing some physical quantity. We will focus on
expansions in the coupling constant $U$ and a shift in the chemical potential $\delta\mu$. 
The calculation of each coefficient involves the evaluation of many Feynman diagrams expressed
in terms of the bare propagator, in our case taken as a function of momentum and two imaginary times. The evaluation of a diagram
then boils down to a sum over multiple momentum variables and a multiple imaginary time integral that is always of the
same generic form. The goal of this Section is to find a general analytical
solution for these time integrals and reformulate the perturbation series as a function of a complex frequency $z$.

\subsection{Analytical solution of time-integrals}
\label{sec:analytical_solution}

We are interested in solving analytically $N-1$-fold integrals over $\{\tau_{i=2..N}\}$ of the form
\begin{equation}\label{eq:integral}
 {\cal I}_\mathbf{X}(i\Omega_\eta) = \prod_{i=2}^{N}
 \int_{0}^{\tau_{i+1}} \mathrm{d}\tau_i\; \tau^{l_i}_i \;
  e^{\tau_{i}(i\Omega_\eta\delta_{r,i}+\omega_i)} 
\end{equation}
where the parameters of the integrand are given by
\begin{equation}
 \mathbf{X}=(r,\{l_2...l_N\}, \{\omega_2...\omega_N\})
\end{equation}
The argument $r$ is integer and determines which of the times $\tau_i$ is
multiplied by the external Matsubara frequency $i\Omega_\eta$ in the
exponential. The frequency $i\Omega_\eta$
can be any Matsubara frequency, either fermionic or bosonic, depending on $\eta$;
$i\Omega_{\eta=-1}\equiv i\omega\equiv i(2m+1)\pi T$ and $i\Omega_{\eta=1}\equiv i\nu\equiv 2im\pi T$, with $m\in {\mathbb{Z}}$.
The integer powers of $\tau_i$ outside of the exponent are given by $l_i\geq0$, and the parameters $\omega_i$ may be complex.
The limit of the outermost integration is the inverse temperature $\tau_{N+1}\equiv\beta$. 
We denote by $\delta_{x,y}$ the Kronecker delta (it will be used throughout this paper, also in the shortened version $\delta_x\equiv\delta_{x,0}$). 
The reason for our choice to label times starting from 2 will become clear later.

The main insight is that upon applying the innermost integral, one gets a number of terms, but each new integrand has the same general form $\sim\tau^n e^{\tau z}$.
The solution therefore boils down to a recursive application of
\begin{equation}\label{eq:int1}
 \int_0^{\tau_\mathrm{f}}  \tau^n e^{\tau z} \mathrm{d}\tau = \sum_{k=0}^{n+1} (-)^k C_{nk}\frac{\tau_\mathrm{f}^{n+1-k-B_{nk}}e^{B_{nk}z\tau_\mathrm{f}}}{z^{k+B_{nk}}} %+ \frac{(-)^{n+1}C_{nn}}{A^{n+1}}
\end{equation}
with $ B_{nk} = 1-\delta_{k,n+1} $ and $C_{nk} = \frac{n!}{(n-k+\delta_{k,n+1})!}$ (for proof see Appendix \ref{sec:proof_eqint1}), and
\begin{equation}\label{eq:int2}
\lim_{z\rightarrow 0}\int_0^{\tau_\mathrm{f}} \tau^n e^{\tau z} \mathrm{d}\tau = \frac{\tau_\mathrm{f}^{n+1}}{n+1}
\end{equation}

The number of terms obtained after each integration is apparently $1+(1-\delta_{z})(n+1)$, and we can enumerate all terms obtained after the full integration by a set of integers $\{k_{i=2..N}\}$,
where $k_i\geq0$ denotes the choice of the term of the integral $i$ (over d$\tau_i$).
%. $\tilde{z}_i$ is the exponent of $e$ in the integration $i$, and likewise, $n_i$ is the exponent of $\tau_i$.

For a given choice of $\{k_i\}$,
the propagation of exponents ($n$ and $z$ in Eqs~\ref{eq:int1} and \ref{eq:int2}) across successive integrals can be fully described by a simple set of auxiliary quantities.
The exponent of $e$ in the integration $i$ we denote as $\tilde{z}_i$ and it is given by
\begin{equation} \label{eq:ztilde_def}
 \tilde{z}_i\equiv z_i+b_{i-1}\tilde{z}_{i-1},\;\;\;\tilde{z}_{2}\equiv z_{2}
\end{equation}
\begin{equation}
 z_i \equiv \delta_{i,r} i\Omega_\eta+\omega_i
\end{equation}
where we introduced 
$
b_i\equiv B_{n_{i},k_{i}} 
$.
The meaning of $b_i$ can be understood by looking at Eq.~\ref{eq:int1}: The exponent of $e$ that enters the integral on the left-hand side survives in all but the last term ($k=n+1$) on the right-hand side. Therefore, $b_i=1$ means that the exponent propagates from integration $i$ to integration $i+1$, while $b_i=0$ means it does not, and the calculation of the recursive $\tilde{z}_i$ is reset with each $b_i=0$.
The auxiliary quantity $n_i$ is the exponents of $\tau_i$ and is specified below.

We will need to obtain a more convenient expression for the exponent $\tilde{z}_i$, where $i\Omega_\eta$ appears explicitly. Straightforwardly, we can write 
\begin{equation}\label{eq:tildez_of_tildeom}
 \tilde{z}_{i} = i\Omega_\eta h_{i}+\tilde{\omega}_{i}
\end{equation}
with auxiliary quantities
\begin{equation}\label{eq:omega_tilde_definition}
 \tilde{\omega}_{i} \equiv \omega_{i}+b_{i-1}\tilde{\omega}_{i-1},\;\;\;\tilde{\omega}_{2}\equiv \omega_{2}
\end{equation}
and
\begin{equation}
 h_{i} \equiv \left\{\begin{array}{cc}
              0, & i<r \\
              1, & i=r \\
              b_{i-1} h_{i-1} & i>r)
            \end{array} \right.
\end{equation}
To be able to determine whether the exponent in the integrand, $\tilde{z}_{i}$, is zero and then employ Eq.\ref{eq:int2} if needed, we can now use
\begin{equation}\label{eq:practical_delta_z}
\delta_{\tilde{z}_{i}} = \left\{ \begin{array}{cc}
		    1, & h_i=0 \wedge \tilde{\omega}_i=0\\
		    0, & \mathrm{otherwise}
                \end{array}\right.
\end{equation}
It is important to note that at the time of integration $i\Omega_\eta$ is unspecified and whether $\tilde{z}_{i}$ is zero cannot be tested by numerical means, unless $i\Omega_\eta$ does not appear in $\tilde{z}_{i}$. With the convenient rewriting of Eq.\ref{eq:ztilde_def} as Eq.\ref{eq:tildez_of_tildeom}, one can tell whether $i\Omega_\eta$ appears in $\tilde{z}_{i}$ by looking at $h_i$. If $i\Omega_\eta$ does appear in $\tilde{z}_{i}$ (i.e. $h_i=1$), we cannot use Eq.\ref{eq:int2} even if one can find such $i\Omega_\eta$ that cancels $\tilde{\omega}_{i}$. This is because we are working towards an analytical expression which ought to be general for all possible $i\Omega_\eta$.

The exponent of $\tau$ that will be carried over from integration $i$ to integration $i+1$ depends on the choice of the term from the integral $i$, and is given by
$\mathrm{Pos}(n_{i}-k_{i})$ where $\mathrm{Pos}$ denotes the positive part of the number ($\mathrm{Pos}(x)=(x+|x|)/2$).
$n_i$ denotes the maximum exponent that can be carried over from integration $i$, and is obtained as:
\begin{equation}
n_{i}=\left\{ \begin{array}{cc}
		    \delta_{\tilde{z}_{i}}+l_i+\mathrm{Pos}(n_{i-1}-k_{i-1}), & i>2 \\
		    \delta_{\tilde{z}_{i}}+l_i, & i=2 
                \end{array}\right.
\end{equation}

In the case of Eq.\ref{eq:int1}, the maximal exponent that can be carried over to the next integration coincides with the exponent that entered the integral (the integral Eq.\ref{eq:int1} does not raise the power of $\tau$), so the definition of $n_i$ coincides with the meaning of $n$ in Eq.\ref{eq:int1}. In the case of integral Eq.\ref{eq:int2}, $n_i$ rather denotes the exponent after the integration, i.e. $n+1$.

\begin{table*}[ht!]
\begin{center}
\begin{tabular}{c|ccc|c|ccccc|l|l||c}
$i$ & $\delta_{r,i}$ & $l_i$ & $\omega_i$ & $k_i$ & $b_i$ & $n_i$ &$\tilde{\omega}_i$& $h_i$ &$\delta_{\tilde{z}_i}$ &        integrand                &         integral & total\\
\hline
2   &    0           &  0    &  1         &  {\red 0}    &  1    &  0    &    1    & 0      &  0 & $e^{\tau_2\cdot 1} $            &  ${\blue \frac{1}{1}}{\red e^{\tau_3\cdot 1}}- \frac{1}{1}1$                                         &\\
3   &    0           &  1    &  2         &  {\red 1}    &  1    &  1    &    3    & 0      &  0 & $\tau_3e^{\tau_3(2+1)} $        &  $\frac{1}{3}\tau_4e^{\tau_4\cdot 3}-{\blue\frac{1}{3^2}}{\red e^{\tau_4\cdot 3}}+ \frac{1}{3^2}1$   &              \\  
4   &    1           &  0    &  1         &  {\red 1}    &  0    &  0    &    4    & 1      &  0 & $e^{\tau_4(i\Omega_\eta+1+3)} $ &  $\frac{1}{i\Omega_\eta+4}e^{\tau_5(i\Omega_\eta+4)}- {\blue \frac{1}{i\Omega_\eta+4}}{\red 1}$      & ${\blue \frac{1}{1}\left(-\frac{1}{3^2}\right)\left(-\frac{1}{i\Omega_\eta+4}\right)\frac{1}{1} \frac{1}{4}}{\red  \beta e^{\beta\cdot 4}}$           \\  
5   &    0           &  0    &  0         &  {\red 0}    &  1    &  1    &    0    & 0      &  1 & $e^{\tau_5\cdot 0} $            &  ${\blue \frac{1}{1}}{\red \tau_6^1}$                                                                &  $\rightarrow \frac{\beta e^{4\beta}/36}{(z-(-4))^1}$\\  
6   &    0           &  0    &  4         &  {\red 0}    &  1    &  1    &    4    & 0      &  0 & $\tau_6 e^{\tau_6\cdot 4} $     &  ${\blue \frac{1}{4}}{\red \beta e^{\beta\cdot 4}}-\frac{1}{4^2}e^{\beta\cdot 4}+ \frac{1}{4^2}1$    &             \\  
\end{tabular}
\end{center}
\caption{Illustration of the calculation of a single term in Eq.\ref{eq:main_analytical_solution}. Rows correspond to successive integrations over $\mathrm{d}\tau_i$. The second to fourth columns are parameters of the integrand. The choice of the term is colored red. The remaining columns are auxiliary quantities, the integrand before and after each integration. The prefactors that are ``collected'' after each integration are written in blue. The full contribution is written in the last column and then simplified to the form of a term in Eq.\ref{eq:analytical_solution_general_form}.}
\label{tab:only}
\end{table*}

After the last integration, it can happen that $i\Omega_\eta$ appears in the exponent of $e$ (this is signaled by $h_Nb_N=1$). We can then use the property $e^{i\Omega_\eta\beta} = (-1)^{\delta_{\eta,-1}}$ to eliminate it from this exponent. Then, the solution for the integral can be continued to the whole of the complex plane $i\Omega_\eta\rightarrow z$, and can be written down as (introducing the additional superscript $\eta$ because the fermionic/bosonic nature of the expression can no longer be inferred from the external Matsubara frequency)
\begin{eqnarray} \label{eq:main_analytical_solution}
  {\cal I}^\eta_\mathbf{X}(z) &=&
  \sum_{\{b_i\in[\delta_{\tilde{z}_i},1]\}_{i=2..N}}
   e^{b_N\beta\tilde{\omega}_{N}}
  \sum_{\{k_i\in[0,(1-\delta_{\tilde{z}_i})n_i]\}_{i:b_i=1}}   
   \\ \nonumber
&& \;\;\;\times      
\prod_{i: \delta_{\tilde{z}_{i}}=1}\;\; \frac{1}{n_{i}} \\ \nonumber
&& \;\;\;\times(-1)^{b_{N}h_{N}\delta_{\eta,-1}+\sum_{i=2}^{N} k_i} \times  \beta^{n_{N}+1-b_{N}-k_{N}}\\ \nonumber
&& \;\;\;\times  
  \prod_{
    i:h_{i}=0 \wedge \tilde{\omega}_{i}\neq0
  }\frac{C_{n_i,k_i}}{\tilde{\omega}_i^{k_i+b_i}} \\ \nonumber
&& \;\;\;\times
  \prod_{
    i:h_{i}=1
  }\frac{C_{n_i,k_i}}{(z+\tilde{\omega}_i)^{k_i+b_i}} 
\end{eqnarray}
Note that we have expressed the sum over $\{k_i\}$ as a sum over $\{b_i\}$ and a partial (inner) sum over $\{k_i\}$. This is not necessary, being that $b_i$ is a function of $k_i$. Each $b_i$ is fully determined by $k_i$, but not the other way around, so the inner sum over $k_i$ in Eq.\ref{eq:main_analytical_solution} goes over values that are allowed by the corresponding $b_i$. We present this form of Eq.\ref{eq:main_analytical_solution} to emphasize that the factor $e^{b_N\beta\tilde{\omega}_{N}}$ depends only on $\{b_i\}$, and can thus be pulled out of the inner $\{k_i\}$ sum. The notation ``$i: b_i=1$'' means that we only consider indices $i$ such that $b_i=1$.  
We therefore only sum over those $k_i$ for which the corresponding $b_i=1$. The remaining $k_i$ are
fixed to $n_i+1$, which is the only possibility if $b_i=0$. 
The notation is applied analogously in other products over $i$. 

The only remaining step is to expand the product of poles in Eq.\ref{eq:main_analytical_solution} into a sum of poles (see Ref.\onlinecite{VucicevicPRB2020} for more details)
\begin{eqnarray} \label{eq:main_transformation}
&&\prod_\gamma \frac{1}{(z-z_\gamma)^{m_\gamma}} = 
  \sum_\gamma \sum_{r=1}^{m_\gamma} \frac{1}{(z-z_\gamma)^{r}} \times \\ \nonumber
&& \;\;\;\;\;\;\;\;\;\; \times(-1)^{m_\gamma-r} \sum_{{\cal C}\{p_{\gamma'\neq\gamma} \in {\mathbb N_0}\}:\sum_{\gamma'\neq\gamma} p_{\gamma'} = m_\gamma-r} \times \\ \nonumber 
&& \;\;\;\;\;\;\;\;\;\; \times \prod_{\gamma'\neq\gamma}\frac{(m_{\gamma'}+p_{\gamma'}-1)!}{p_{\gamma'}!(m_{\gamma'}-1)!} \frac{1}{(z_\gamma-z_{\gamma'})^{m_{\gamma'}+p_{\gamma'}}}
\end{eqnarray}
and the final expression has the form
\begin{equation}\label{eq:analytical_solution_general_form}
 {\cal I}^\eta_\mathbf{X}(z) = \sum_{j,p\in\mathbb{N}} \frac{{\cal A}_{j,p}}{(z-{\cal Z}_j)^p}
\end{equation}

In order to illustrate our solution, we present in tabular form (Table~\ref{tab:only}) a summary of all intermediate steps, integrand parameters and auxiliary quantities that are used in calculating the contribution for a single choice of $\{k_i\}$, in an example with $N=6$ and $r=4$.

Also note that if $r\notin [2,N]$ (no Matsubara frequency appearing in any exponent), the result of the integral is a number, rather than a frequency dependent quantity. In that case, the integral can be straightforwardly generalized to the case of real time, where integrations go to some externally given time $t$ (instead of $\beta$), and the resulting expression is a function of that time. The step Eq.\ref{eq:main_transformation} is then not needed. See Appendix \ref{sec:real_time} for details.

\subsubsection{Numerical evaluation of the analytical expression and relation to other algorithms}

The implementation of Eq.\ref{eq:main_analytical_solution} is rather
straightforward and much simpler than the algorithmic Matsubara summations in
our previous work Ref.~\onlinecite{VucicevicPRB2020}. Indeed, most of the
calculations just require the numerical evaluation of an analytical expression
and it is not necessary to implement a dedicated symbolic algebra to manipulate
the expressions. The only exception is the last step
Eq.\ref{eq:main_transformation}. This transformation was the centerpiece of the
algorithm in Ref.~\onlinecite{VucicevicPRB2020} and was applied recursively
many times, leading to complex book-keeping and data structures. 
Ultimately, the result was a symbolic expression that was stored,
and a separate implementation was needed for the comprehension and numerical evaluation of such a general symbolic expression.
In the present context, however, Eq.\ref{eq:main_transformation} is applied only once to produce numbers,
and is simple to implement.

The other important point is that we treat analytically cases with
$\delta_{\tilde{z}_i}=1$ by employing Eq.\ref{eq:int2}. With the
frequency-summation algorithms\cite{VucicevicPRB2020,taheridehkordiPRB2020},
one cannot take into account possible cancellations of $\omega_i$ terms in Eq.\ref{eq:omega_tilde_definition}
without computing a large number of
separate analytical solutions. When untreated, these cancellations yield
diverging ratios in the final expressions, which need to be regularized. On the
contrary, in Eq.\ref{eq:main_analytical_solution} the ratio
$1/\tilde{\omega}_i^{k_i+b_i}$ cannot have a vanishing denominator and its
size will in practice be limited by the energy resolution. This will also allow
us to have the final result in the form of a sum of poles on an equidistant
grid on the real-axis, and extract the real-axis results without any numerical
pole-broadening (see Section~\ref{sec:numerical_evaluation} and Appendix~\ref{sec:broadening}).

It is interesting to compare the computational effort for the numerical evaluation of our analytical solution to the straightforward numerical integration.
In the most straightforward integration algorithm, one would discretize the imaginary-time interval $[0,\beta]$ with $N_\tau$ times, and then perform the summation which has the complexity $O(N_\tau^{N-1})$ for each external $\tau$, so overall $O(N_\tau^{N})$.
%While our algorithm is indeed very efficient at low orders (more efficient than our implementation of the Matsubara summation algorithm\cite{VucicevicPRB2020} up to order 6), 
With our algorithm we do not have to go through all configurations of internal times, but we do need to go through all possible permutations of the internal times, and for each permutation there is at least $2^{N-1}$ terms to be summed over.
So the number of terms one has to sum grows at least as $O((N-1)! 2^{N-1})$. At sufficiently high $N$, this number is bound to outgrow the exponential $N_\tau^{N}$, whatever the $N_\tau$. This will happen, however, only at very large $N$. For example, if $N_\tau=30$, the analytical solution becomes slower at around $N=40$. Moreover, one actually needs a much larger $N_\tau$, especially at low temperature. In any case, the additional computational effort can be understood as coming from the difference in the information content of the result, which is a lot more substantial in the case of the analytical solution.

At orders $N<6$ (within context of DiagMC), we find that the implementation of our new algorithm is significantly more efficient than our current implementation of the Matsubara summations from Ref.\onlinecite{VucicevicPRB2020}, and at $N=6$ they are about equally efficient. However, we anticipate that further optimizations will be possible at the level of Eq.\ref{eq:main_analytical_solution}.

\subsection{Expansion of the bare propagator} \label{sec:g0_expansion}

The central quantity is the Green's function defined in Matsubara formalism as:
\begin{eqnarray}
 G_{\sigma\mathbf{k}}(\tau-\tau') &=& -\langle T_\tau c_{\sigma\mathbf{k}}(\tau)c^\dagger_{\sigma\mathbf{k}}(\tau')\rangle \\ \nonumber
 &=&
 \left\{\begin{array}{cc} 
   -\langle c_{\sigma\mathbf{k}}(\tau)c^\dagger_{\sigma\mathbf{k}}(\tau')\rangle, & \tau>\tau' \\
   \langle c^\dagger_{\sigma\mathbf{k}}(\tau') c_{\sigma\mathbf{k}}(\tau)\rangle, & \tau'>\tau   
  \end{array}\right.
\end{eqnarray}
where $\tau,\tau' \in [0,\beta]$.
The non-interacting Green's function (or the bare propagator) in the eigenbasis of the non-interacting Hamiltonian has a very simple general form
\begin{equation}\label{eq:g0_basic}
 G_{0}(\varepsilon,i\omega) \equiv \frac{1}{i\omega-\varepsilon}
\end{equation}
and for the plane-wave $\mathbf{k}$, the propagator is $G_{0,\mathbf{k}}(i\omega) = G_{0}(\varepsilon_\mathbf{k}-\mu,i\omega)$.

As we will discuss below, the diagrammatic series for the self-energy will in general be constructed
from different powers of the bare propagator:
\begin{equation}
 G_{0}^l(\varepsilon,i\omega) \equiv \frac{1}{(i\omega-\varepsilon)^l}
\end{equation}
Indeed, these powers naturally arise after expanding the bare propagator in a Maclaurin series 
$
 \frac{1}{z+x}= \sum_{n=0}^{\infty} \frac{(-x)^n}{z^{n+1}}
$
around a small chemical potential shift
\begin{eqnarray} \label{eq:g0_expansion}
 G_0(\varepsilon,i\omega)  
  &=& \sum_{l=1}^{\infty} (-\delta\mu)^{l-1} G^{l}_{0}(\varepsilon+\delta\mu,i\omega) 
\end{eqnarray}
This series converges (for all $i\omega$) if $\delta\mu$ is smaller in amplitude than the first Matsubara frequency: $|\delta\mu|<\pi T$. Nevertheless, this expression will become a part of a larger series with additional expansion parameters, which may result in a modified convergence radius of the overall series with respect to $\delta\mu$.

We anticipate that the Feynman diagrams will be formulated in the imaginary-time domain, so it is essential to work out the Fourier transform of $G_{0}^l(\varepsilon,i\omega)$. We present the full derivation in Appendix~\ref{sec:g0tau} and here only write the final solution
\begin{eqnarray}\label{eq:g0_tau}
 &&G^{l}_{0}(\varepsilon, \tau-\tau') = \\ \nonumber
 &&\;\;\;\;\;
   s_{\tau,\tau'}e^{-\varepsilon(\tau-\tau')}n_\mathrm{F}(s_{\tau,\tau'}\varepsilon)\;
   \sum_{\zeta=0}^{l-1}\sum_{\varsigma=0}^{l-\zeta-1}
   %\sum_{\zeta,\varsigma=0}^{l} 
   c^{s_{\tau,\tau'}}_{l\zeta\varsigma}(\varepsilon) \; \tau^\zeta {\tau'}^\varsigma
\end{eqnarray}
with $s_{\tau,\tau'}=\mathrm{sgn}(\tau'-\tau)$. In our notation, $l$ in $G_0^l$ is a superscript index, rather than the power of $G_0$ (although these meanings coincide in the case of $G_0^l(\varepsilon,i\omega)$).
The Fermi function is defined as
$
 n_\mathrm{F}(\varepsilon)=1/(e^{\beta\varepsilon}+1)
$
and the coefficients that go with $\tau^\zeta \tau'^\varsigma$ terms are
\begin{eqnarray}\nonumber
    c_{l,\zeta,\varsigma}^{-}(\varepsilon)&=&
    \sum_{n=0}^{l-\varsigma-\zeta-1}\frac{n!(-1)^{l+\varsigma-1}\left(-n_{\mathrm{F}}(\varepsilon)\right)^{n}
    \beta^{l-\varsigma-\zeta-1}}{(l-\varsigma-\zeta-1)!(\varsigma+\zeta)!} \times \\ \label{eq:coefficients}
    && \;\;\;\;\;\;\;\;\times
    {l-\varsigma-\zeta-1\brace n}
    \binom{\varsigma+\zeta}{\zeta}
\end{eqnarray}
and $c_{l,\zeta,\varsigma}^{+}(\varepsilon)=(-1)^{l-1}c_{l,\varsigma,\zeta}^{-}(-\varepsilon) $.
Here we make use of binomial coefficients
$
 \binom{n}{k} = \frac{n!}{k!(n-k)!}
$
and the Stirling number of the second kind
$
 {n \brace k } = \sum_{i=0}^{k} \frac{(-1)^i}{k!} \binom{k}{i} (k-i)^n
$.

\subsection{Application to DiagMC}\label{sec:application_to_diagmc}

In the following, we apply the analytic time-integral and the expansion of the bare propagator in the context of DiagMC.
We discuss two kinds of self-energy series (Hartree shifted and bare) and the corresponding implementation details.
Note that some symbols will be redefined with respect to previous sections.

\subsubsection{Hartree-shifted series}

In this section we discuss the construction of the self-energy series, where all tadpole-like insertions are omitted in the topologies of the diagrams. Rather, the full Hartree shift is absorbed in the bare propagator. The diagrams are therefore expressed in terms of the Hartree-shifted bare propagator
\begin{equation}
 G^{\mathrm{HF}}_{0,\mathbf{k}}(i\omega) = G_0(\tilde{\varepsilon}_\mathbf{k}, i\omega)
\end{equation}
with the Hartree-shifted dispersion defined as
\begin{equation}
\tilde{\varepsilon}_\mathbf{k} = \varepsilon_\mathbf{k} - \mu +U\langle n_{\sigma}\rangle 
\end{equation}
where $\langle n_{\bar\sigma} \rangle$ is the average site-occupation per spin.

\begin{figure*}[ht!]
 \begin{center}
 \includegraphics[width=5.8in, trim=0cm 0cm 0cm 0cm, clip]{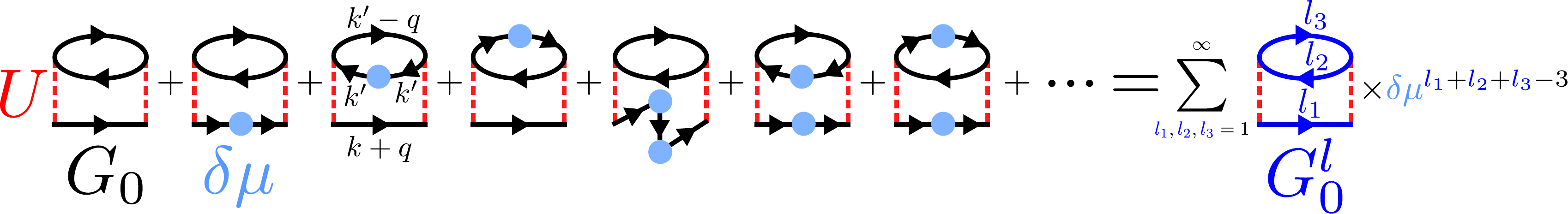}
  \end{center}
 \caption{Illustration of the use of $G_0^l(\varepsilon,\tau-\tau')$ propagator. The entire series of diagrams with all possible $\delta\mu$ insertions can be captured by a single diagram with additional degrees of freedom.
 }
 \label{fig:g0l_diags}
\end{figure*}

After constructing the tadpole-less topologies, we are free to expand all propagators that appear in the diagrams according to Eq.\ref{eq:g0_expansion}:
\begin{eqnarray} 
 G^{\mathrm{HF}}_{0,\mathbf{k}}(i\omega)  
  &=& \sum_{l=1}^{\infty} (-\delta\mu)^{l-1} G^{l}_{0}(\tilde{\varepsilon}_\mathbf{k}+\delta\mu,i\omega) 
\end{eqnarray}
In frequency domain, this step can be viewed as introducing new topologies: we now have diagrams with any number of single-particle-vertex ($\delta\mu$) insertions on any of the propagator lines. Each arrangement of these additional single-particle vertices on the diagram does require a separate solution by the symbolic algebra algorithm as presented in Refs.~\onlinecite{taheridehkordiPRB2020,VucicevicPRB2020}. Nevertheless, as a $\delta\mu$-vertex cannot carry any momentum or energy, the formal effect of it is that it just raises the power $l$ of the propagator that passes through it. In the imaginary-time domain, it turns out that the contribution of $\delta\mu$-dressed diagrams is readily treatable by the analytical expression Eq.\ref{eq:main_analytical_solution}, and we no longer have to view the $\delta\mu$-insertions as changes to topology, but rather as additional internal degrees of freedom to be summed over. This is illustrated in Fig.\ref{fig:g0l_diags}.

Up to the Hartree shift, the self-energy expansion can now be made in powers of the interaction $U$ and the small chemical-potential shift $\delta\mu$
\begin{eqnarray}
 &&\Sigma^{(\mathrm{HF})}_{\mathbf{k}}(\tau) 
 = \sum_N (-U)^N \times \\ \nonumber
 &&\;\;\;\;\;\;\;\;\;\;\;\;\;\;\times  \sum_{l_1,...,l_{2N-1}=1}^{\infty} (-\delta\mu)^{\sum_j (l_j-1)} \sum_{\Upsilon_N} D_{\Upsilon_N,\mathbf{k},\{l_j\},\delta\mu}(\tau) 
\end{eqnarray}
where $j$ enumerates the propagators, of which there are $N_\mathrm{prop} = 2N-1$, $N$ is perturbation order in $U$, each $l_j$ goes from 1 to $\infty$, $\Upsilon_N$ enumerates distinct topologies of the diagram at order $N$ (without any $\delta\mu$ or Hartree insertions), and $D$ is the contribution of the diagram. The general form of the diagram contribution is
\begin{eqnarray} \label{eq:D_starting_point}
&&D_{\Upsilon_N,\mathbf{k},\{l_j\},\delta\mu}(\tau) =\\ \nonumber
&&\;\;\;\;(-1)^{N_\mathrm{bub}}\prod_{i=2}^{N-1} \int_0^\beta \mathrm{d}\tau_i \sum_{\mathbf{k}_1..\mathbf{k}_N}  \prod_{j=1}^{2N-1} 
G^{l_j}_{0}(\bar{\varepsilon}_{\tilde{\mathbf{k}}_j},\tilde{\tau}_{j}-\tilde{\tau}_{j}')
\end{eqnarray}
with $\bar{\varepsilon}_{\mathbf{k}}\equiv\tilde{\varepsilon}_{\mathbf{k}}+\delta\mu$. 
We denote $N_\mathrm{bub}$ the number of closed fermion loops in the diagram;  $\tau_1...\tau_{N-1}$ are internal times, and we fix $\tau_{i=1}=0$;
$\tau$ is the external time; $\mathbf{k}$ is the external momentum, $\mathbf{k}_1..\mathbf{k}_N$ are the independent internal momenta;
$j$ indexes the propagator lines, and $\tilde{\mathbf{k}}$ are corresponding linear combinations of the momenta 
$\tilde{\mathbf{k}}_j\equiv\sum_{\lambda=0}^{N} \tilde{s}_{j\lambda} \mathbf{k}_\lambda$, 
where $\tilde{s}_{j\lambda}\in\{-1,0,1\}$ and we index with 0 the external momentum 
$\mathbf{k}_0\equiv\mathbf{k}$. 
$\tilde{\tau}_j$ and $\tilde{\tau}'_j$ are outgoing and incoming times for the propagator $j$, and take values in $\{ \tau_1 ... \tau_N \}$, where we denote with index $N$ the external time $\tau_N\equiv\tau$. The coefficients $\tilde{s}_{j\lambda}$, times $\tilde{\tau}_j,\tilde{\tau}'_j$ and the number $N_\mathrm{bub}$ are implicit functions of the topology $\Upsilon_N$. Throughout the paper, we assume normalized $\mathbf{k}$-sums, $\sum_\mathbf{k}\equiv \frac{1}{N_\mathbf{k}}\sum_\mathbf{k}$, where $N_\mathbf{k}$ is the number of lattice sites.

We can perform the Fourier transform of the external time, to obtain the contribution of the diagram in the Matsubara-frequency domain:
\begin{eqnarray}
&& D_{\Upsilon_N,\mathbf{k},\{l_j\},\delta\mu}(i\omega) = \\ \nonumber
&& \;\;\;(-1)^{N_\mathrm{bub}}\prod_{i=2}^{N} \int_0^\beta \mathrm{d}\tau_i e^{i\omega\tau_N} \sum_{\mathbf{k}_1..\mathbf{k}_N} \prod_{j=1}^{2N-1}  G^{l_j}_{0}(\bar{\varepsilon}_{{\mathbf{k}}_j},\tilde{\tau}_{j}-\tilde{\tau}_{j}')
\end{eqnarray}

The Green's function $G^l_0(\varepsilon,\tau-\tau')$ is discontinuous at $\tau=\tau'$, so to be able to perform the $\tau$-integrations analytically, we first need to split the integrals into ordered parts
\begin{eqnarray}
 && \int_0^\beta \mathrm{d}\tau_2 ...\int_0^\beta \mathrm{d}\tau_N = \sum_{(\tau_{p_2}..\tau_{p_N})\in{\cal P}(\{\tau_2..\tau_N\})} \times \\ \nonumber
 && \;\;\times \int_0^\beta \mathrm{d}\tau_{p_N} \int_0^{\tau_{p_N}} \mathrm{d}\tau_{p_{N-1}}  ... \int_0^{\tau_{p_4}} \mathrm{d}\tau_{p_3} \int_0^{\tau_{p_3}} \mathrm{d}\tau_{p_2}
\end{eqnarray}
where $\cal P$ denotes all $(N-1)!$ permutations of time indices. $p$ labels the permutation and $p_i$ is the permuted index
of vertex $i$.

Let us rewrite the contribution of the diagram, with propagators written explicitly using the expression Eq.\ref{eq:g0_tau}:
\begin{eqnarray}
&& D_{\Upsilon_N,\mathbf{k},\{l_j\},\delta\mu}(i\omega) = (-1)^{N_\mathrm{bub}} \sum_{\mathbf{k}_1..\mathbf{k}_N} \\ \nonumber
&& \;\;\times 
\sum_{(\tau_{p_2}..\tau_{p_N})\in{\cal P}(\{\tau_2..\tau_N\})} (-1)^{N_{\mathrm{fwd}}(p)} \prod_{j} n_F(s_j\bar{\varepsilon}_{\tilde{\mathbf{k}}_j}) \\ \nonumber
&& \;\;\times 
\sum_{\zeta_j=0}^{l_j-1}\sum_{\varsigma_j=0}^{l_j-\zeta_j-1} c_{l_j,\zeta_j,\varsigma_j}^{s_j}(\bar{\varepsilon}_{\tilde{\mathbf{k}}_j})
\prod_{j\in {\cal J}_{\mathrm{i}}(i=1)}\delta_{\zeta_j}\prod_{j\in {\cal J}_{\mathrm{o}}(i=1)}\delta_{\varsigma_j}
\\ \nonumber
&& \;\;\times
  \int_0^\beta \mathrm{d}\tau_{p_N} \int_0^{\tau_{p_N}} \mathrm{d}\tau_{p_{N-1}}  ... \int_0^{\tau_{p_4}} \mathrm{d}\tau_{p_3} \int_0^{\tau_{p_3}} \mathrm{d}\tau_{p_2} e^{i\omega\tau_N}  \\ \nonumber
&& \;\;\times
  \prod_{i=2}^{N} \tau_i^{\sum_{j \in {\cal J}_{\mathrm{i}}(i)} \zeta_j + \sum_{j \in {\cal J}_{\mathrm{o}}(i)} \varsigma_j }
  e^{\tau_i(\sum_{j \in {\cal J}_{\mathrm{o}}(i)}\bar{\varepsilon}_{\tilde{\mathbf{k}}_j} - \sum_{j \in {\cal J}_{\mathrm{i}}(i)}\bar{\varepsilon}_{\tilde{\mathbf{k}}_j})}  
\end{eqnarray}
where ${\cal J}_{\mathrm{i}/\mathrm{o}}(i)$ is the set of incoming/outgoing propagators $j$ of the vertex $i$, which depends on the topology $\Upsilon_N$. 
We also introduced  shorthand notation $s_j=s_{\tilde{\tau}_j,\tilde{\tau}_j'}$. Practically, $s_j$ depends on whether $p(i(j))>p(i'(j)))$ or the other way around, where $i(j)$/$i'(j)$ is the outgoing/incoming vertex of propagator $j$ in the given permutation $p$.
The total number of forward-facing propagators is $N_\mathrm{fwd}(p)=\sum_j \delta_{-1,s_j}$, which depends on the permutation and the topology.
The products of $\delta_{\zeta_j}$ and $\delta_{\varsigma_j}$ are there to ensure that the time $\tau_1=0$ is not raised to any power other than 0, as such terms do not contribute.

Now we can apply the analytic solution for the time integrals (Eq.\ref{eq:main_analytical_solution}) to arrive at the final expression:
\begin{widetext}
\begin{eqnarray} \label{eq:final_expression}
&& D_{\Upsilon_N,\mathbf{k},L,\delta\mu}(z) = (-1)^{N_\mathrm{bub}} \sum_{\{\tilde{l}_j\geq0\}:\sum_j \tilde{l}_j=L}\sum_{\mathbf{k}_1..\mathbf{k}_N}
\sum_{(\tau_{p_2}..\tau_{p_N})\in{\cal P}(\{\tau_2..\tau_N\})} (-1)^{N_{\mathrm{fwd}}(p)}\\ \nonumber
&& \;\;\times 
 \prod_{j} n_F(s_j\bar{\varepsilon}_{\tilde{\mathbf{k}}_j}) 
\sum_{\zeta_j=0}^{\tilde{l}_j}\sum_{\varsigma_j=0}^{\tilde{l}_j-\zeta_j} c_{\tilde{l}_j+1,\zeta_j,\varsigma_j}^{s_j}(\bar{\varepsilon}_{\tilde{\mathbf{k}}_j})
\prod_{j\in {\cal J}_{\mathrm{i}}(i=1)}\delta_{\zeta_j}\prod_{j\in {\cal J}_{\mathrm{o}}(i=1)}\delta_{\varsigma_j} 
{\cal I}^{\eta=-1}_{\mathbf{X}}(z)\\ \nonumber \\\nonumber
&& \mathbf{X}=
\left(p(N),
  \left\{\sum_{j \in {\cal J}_{\mathrm{i}}(i(p_{i'}))} \zeta_j + \sum_{j \in {\cal J}_{\mathrm{o}}(i(p_{i'}))} \varsigma_j 
  \right\}_{i'={2..N}},
  \left\{ \sum_{j \in {\cal J}_{\mathrm{o}}(i(p_{i'}))}\bar{\varepsilon}_{\tilde{\mathbf{k}}_j} - 
     \sum_{j \in {\cal J}_{\mathrm{i}}(i(p_{i'}))}\bar{\varepsilon}_{\tilde{\mathbf{k}}_j}
  \right\}_{i'={2..N}}
  \right)
\end{eqnarray}
\end{widetext}
where $i(p_{i'})$ is the vertex index $i$ of the permuted index $p_{i'}$ and
we have introduced a new expansion variable $L=\sum_j (l_j-1)$ and a convenient variable $\tilde{l}_j = l_j-1$, so that
\begin{eqnarray}
 &&\Sigma^{(\mathrm{HF})}_{\mathbf{k}}(z) = \\ \nonumber
 &&\;\;\;\; \sum_{K=2}^{\infty} \sum_{N=2}^{K} \sum_{L=0}^{K-N} (-U)^N (-\delta\mu)^L \sum_{\Upsilon_N} D_{\Upsilon_N,\mathbf{k},L,\delta\mu}(z) 
\end{eqnarray}
which is the series we implement and use in practice. The meaning of $K$ is the number of all independent (internal and external) times in the diagram. Note that in ${\cal D}$ we perform only $N-1$ integrations over time. Those are the times associated with $N$ interactions vertices, minus the one that is fixed to zero. The integrations of times associated with $\delta\mu$-insertions have already been performed in Eq.\ref{eq:g0_tau}, and there are $L$ such integrals. Overall, the number of independent times is $K=N+L$. Ultimately, we group contributions by the expansion order $K$, and look for convergence with respect to this parameter.

\subsubsection{Numerical implementation of DiagMC and relation to other algorithms}\label{sec:numerical_evaluation}

The expression Eq.\ref{eq:final_expression} is very convenient for numerical evaluation.

First, we restrict the values of $\bar\varepsilon_\mathbf{k}$ to a uniform grid on the real-axis with the step $\Delta\omega$ ($\bar\varepsilon_\mathbf{k}=j\Delta\omega$). These appear in $\omega_2,...,\omega_K$ as terms with integer coefficients, which means that $\{\omega_i\}$ entering ${\cal I}_\mathbf{X}$ will also be restricted to the same uniform grid. The final result therefore has the form:
\begin{equation}\label{eq:sum_of_poles}
 D_{\Upsilon_N,\mathbf{k},L,\delta\mu}(z) = \sum_{j\in\mathbb{Z},p\in\mathbb{N}} \frac{{\cal A}_{j,p}}{(z-j\Delta\omega)^p}
\end{equation}
This form allows us to reinterpret the finite-lattice results as that of the thermodynamic limit and extract $D_{\Upsilon_N,\mathbf{k},L,\delta\mu}(\omega+i0^+)$ without any numerical broadening (see Appendix \ref{sec:broadening} for details).

In our present implementation we perform a flat-weight (uniform) MC sampling over internal momenta $\{\mathbf{k}_i\}$, and do a full summation of all the other sums, and accumulate the amplitudes ${\cal A}_{j,p}$. There are, however, other options. For example, one may sample $\{\mathbf{k}_i\},\{p_i\},\{b_i\}$ and use $P\equiv\prod_j n_F(s_j\bar{\varepsilon}_{\tilde{\mathbf{k}}_j}) e^{b_N\beta\tilde{\omega}_{N}}$ as the weighting function.
%(variables $\{b_i\}$ and the factor $e^{b_N\beta\tilde{\omega}_{N}}$ feature in ${\cal I}_{\mathbf{X}}$, see Eq.\ref{eq:main_analytical_solution}).
We have checked thoroughly that the factor $P$ correlates closely with the contribution to ${\cal A}_{j,p}$ coming from a given choice of $\{\mathbf{k}_i\},\{p_i\},\{b_i\}$ variables (with other variables summed over), and thus $P$ could be a good choice for a weighting function. However, this requires additional operations related to move proposals and trials, and we have not yet been able to make such an algorithm more efficient than the flat-weight MC. Nevertheless, it is apparent that our approach offers more flexibility than the algorithmic Matsubara summations (AMS). In AMS no convenient weighting function can be defined for the Monte Carlo, so one either does the flat-weight summation~\cite{VucicevicPRB2020} or uses the whole contribution to the result as the weight, which comes at the price of having to repeat the calculation for each frequency of interest~\cite{taheridehkordiPRB2020} (on the contrary, in Ref.~\onlinecite{VucicevicPRB2020}, as well as in this paper, the entire frequency dependence of self-energy is obtained in a single MC run). At present it is unclear which scheme is best - whether one should evaluate $D(z)$ one $z$ at a time or capture all $z$ at once as we do here. This choice, as well as the choice of the weighting function, likely needs to be made on a case-by-case basis, as it is probable that in different regimes, different approaches will be optimal. In that sense, the added flexibility of our time-integration approach in terms of the choice of the weighting function may prove valuable in the future.

Concerning floating point arithmetic, it is important that the factor $e^{b_N\beta\tilde{\omega}_{N}}$ stemming from ${\cal I}_{\mathbf{X}}$ can always be absorbed into the product of $n_\mathrm{F}$ functions in the second row of Eq.\ref{eq:final_expression}. This can be understood as follows. A given $\bar{\varepsilon}_{\tilde{\mathbf{k}}_j}$ can at most appear twice as a term in $\tilde{\omega}_{N}$, once with sign $+1$ and once with sign $-1$, corresponding to the incoming $\tilde{\tau}'_j$ and outgoing $\tilde{\tau}_j$ ends of the propagator $j$. In that case, the exponent cancels. The other possibility is that it appears only once, in which case it must correspond to the later time in the given permutation. If the later time is the outgoing end of the propagator, then the propagator is forward facing, the sign in front is $s=-1$; if it is the incoming end, then the propagator is backward facing, and the sign in front is $s=1$. In both cases we can make use of
\begin{equation}
 e^{s\beta\varepsilon}n_F(s\varepsilon) = n_F(-s\varepsilon)
\end{equation}
Therefore, no exponentials will appear in the final expression. A product of $n_F$ functions is at most 1, and the coefficients $c$ are not particularly big.
Then, the size of the pole amplitudes that come out of Eq.\ref{eq:main_analytical_solution} is determined by the energy resolution ($1/\Delta\omega$) and temperature ($\beta^{n_{N}+1-b_{N}-k_{N}}$). In our calculations so far, the amplitudes remain relatively small. Our approach ensures that we do not have very large canceling terms, like we had in Ref.~\onlinecite{VucicevicPRB2020}. Indeed, we have successfully implemented Eq.\ref{eq:final_expression} without the need for multiple-precision floating-point types.

Compared to the Matsubara frequency summation algorithm\cite{taheridehkordiPRB2019,VucicevicPRB2020, taheridehkordiPRB2020}, Eq.\ref{eq:final_expression} presents an improved generality. Eq.\ref{eq:final_expression} is valid for any number and arrangement of instantaneous (i.e. frequency independent) insertions, i.e. any choice of $\{ \tilde{l}_j \}$. In contrast, the algorithmic Matsubara summation has to be performed for each choice of $\{ \tilde{l}_j \}$ independently, and the resulting symbolic expressions need to be stored. For example, at $N=4$ we have 12 $\Upsilon_N$-topologies. Therefore, at $L=0$ the number of analytical solutions to prepare is 12. However, at $L=2$, this number is 336, i.e. 28-fold bigger (we can place $L=2$ insertions on $2N-1=7$ fermionic lines in $7\cdot 6/2+7 =28$ ways, times 12 $\Upsilon_N$-topologies, i.e. 336).

\subsubsection{Bare series}

We are also interested in constructing a bare series where tadpole insertions
are present in diagram topologies.  Tadpole (or Hartree) insertions are instantaneous
and an evaluation of their amplitudes can be done relatively simply by various
means. At the level of the Hubbard model, the Hartree insertions factor out:
For each Hartree diagram, the internal momentum summations and time-integrations can 
be performed beforehand and only once, leading to a significant speed up.

In the expression Eq.\ref{eq:final_expression}, there is no difference between a Hartree insertion and a chemical-potential vertex insertion.
Therefore, the inclusion of the Hartree insertions can be entirely accounted for in the resummation of the $D_{\Upsilon_N,\mathbf{k},L,\delta\mu}(z)$ contributions from the previous section, with the replacement
\begin{equation}
\bar{\varepsilon}_\mathbf{k}\equiv\varepsilon_\mathbf{k}-\mu+\delta\mu 
\end{equation}
(i.e. full Hartree shift excluded).

%We will redefine the meaning of the expansion parameter $K$ to be the total number of Hartree and $\delta\mu$ insertions, and consider the series only up to $K=5$. 
Note that the expansion of the propagators in $\delta\mu$ is performed in Hartree insertions as well, so we need to account for possible additional $\delta\mu$ insertions inside the Hartree diagrams.
As before, our expansion order will be $K$, which is the total number of independent times, with each time associated to a single interaction or a $\delta\mu$-vertex,
including those within Hartree insertions.

\begin{figure}[t!]
 \begin{center}
 \includegraphics[width=3.2in, trim=0.cm 0cm 0cm 0.0cm]{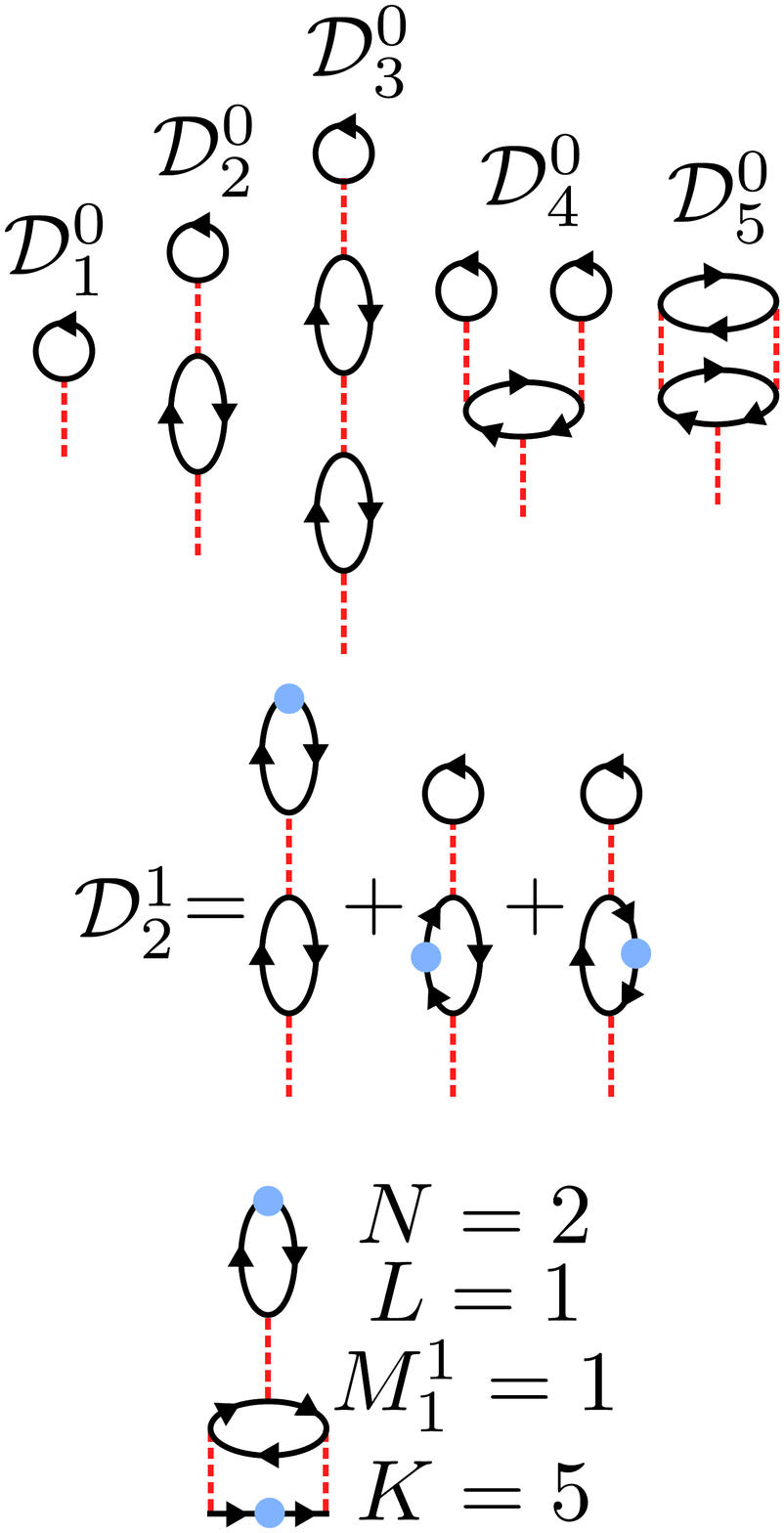}%_horizontal}
  \end{center}
 \caption{Top: Illustration of possible Hartree diagrams, without any $\delta\mu$-insertions. Middle: amplitude of a Hartree diagram with a single $\delta\mu$-insertion. Bottom: an example of a diagram dressed with both Hartree and $\delta\mu$ insertions, and the values of parameters $N,L,\{M_i^{L'}\},K$ it falls under (with $M_{i\neq 1}^{L'\neq 1}=0$).
 }
 \label{fig:tadpoles}
\end{figure}

We will for now focus on the series up to $K=5$. As the number of interactions in $\Upsilon_N$ is at least 2, we can have at most three interaction vertices in a Hartree insertion. There are only 5 such Hartree diagrams (Fig.\ref{fig:tadpoles}). We can evaluate these 5 amplitudes with very little effort, by making use of spatial and temporal Fourier transforms. 

% \begin{figure*}[t!]
%  \begin{center}
%  \includegraphics[width=5.5in, trim=0.cm 0cm 0cm 0.0cm]{Hartree_diagrams_illustration_horizontal}
%   \end{center}
%  \caption{Top: Illustration of possible Hartree diagrams, without any $\delta\mu$-insertions. Middle: amplitude of a Hartree diagram with a single $\delta\mu$-insertion. Bottom: an example of a diagram dressed with both Hartree and $\delta\mu$ insertions, and the values of parameters $N,L,\{M_i^{L'}\},K$ it falls under (with $M_{i\neq 1}^{L'\neq 1}=0$).
%  }
%  \label{fig:tadpoles}
% \end{figure*}

Before we proceed with the calculation of the amplitudes ${\cal D}$ of possible Hartree insertions relevant for the series up to $K=5$, we define some auxiliary quantities.
We first define the bare density
\begin{equation}
 n_0^{\tilde{l}} = \sum_\mathbf{k} G^{l=1+\tilde{l}}_{0}(\bar{\varepsilon}_\mathbf{k},\tau=0^-)
\end{equation}
and the real-space propagator:
\begin{equation}
 G^{l=1+\tilde{l}}_{0,\mathbf{r}} = \sum_\mathbf{k} e^{i\mathbf{k}\cdot\mathbf{r}} G^{l=1+\tilde{l}}_{0}(\bar{\varepsilon}_\mathbf{k},\tau=0^-)
\end{equation}
We will also need the polarization bubble diagram
\begin{eqnarray}
 \chi_{0,\mathbf{r}}^{\tilde{l}_1,\tilde{l}_2}(\tau) = G^{l=1+\tilde{l}_1}_{0,\mathbf{r}}(\tau)G^{l=1+\tilde{l}_2}_{0,-\mathbf{r}}(-\tau) \\
 \chi_{0,\mathbf{q}=0}^{\tilde{l}_1,\tilde{l}_2}(i\nu=0) = \sum_\mathbf{r} \int \mathrm{d}\tau \chi_{0,\mathbf{r}}^{\tilde{l}_1,\tilde{l}_2}(\tau)
\end{eqnarray}
and the second-order self-energy diagram (up to the constant prefactor)
\begin{equation}
 \Sigma_{2,\mathbf{r}}^{\tilde{l}_1,\tilde{l}_2,\tilde{l}_3}(\tau) = G^{l=1+\tilde{l}_1}_{0,\mathbf{r}}(\tau)\chi_{0,\mathbf{r}}^{\tilde{l}_2,\tilde{l}_3}(\tau)
\end{equation}
which can be Fourier transformed to yield $\Sigma_{2,\mathbf{k}}^{\tilde{l}_1,\tilde{l}_2,\tilde{l}_3}(i\omega)$.

We can now calculate the amplitudes of the possible Hartree insertions with a number $L$ of $\delta\mu$ insertions on them, in any arrangement
\begin{eqnarray}\label{eq:hartree_diagrams}
 {\cal D}^{L}_1 &=& (-)n^{L}_0 \\
 {\cal D}^{L}_2 &=& (-)^2\sum_{\substack{ \tilde{l}_1,\tilde{l}_2,\tilde{l}_3  \\ \tilde{l}_1+\tilde{l}_2+\tilde{l}_3=L} } n_0^{\tilde{l}_1} \chi^{\tilde{l}_2,\tilde{l}_3}_{0,\mathbf{q}=0}(i\nu=0) \\ \nonumber
 {\cal D}^{L}_3 &=& (-)^3 \sum_{\substack{ \tilde{l}_1,...,\tilde{l}_5  \\ \sum_i \tilde{l}_i =L} } n^{\tilde{l}_1}_0 \chi^{\tilde{l}_2,\tilde{l}_3}_{0,\mathbf{q}=0}(i\nu=0)\chi^{\tilde{l}_4,\tilde{l}_5}_{0,\mathbf{q}=0}(i\nu=0)\\ \\
 {\cal D}^{L}_4 &=& (-)^3 \sum_{\substack{ \tilde{l}_1,...,\tilde{l}_3  \\ \sum_i \tilde{l}_i =L} } \binom{ 2+\tilde{l}_3}{2} n^{\tilde{l}_1}_0 n^{\tilde{l}_2}_0 n^{2+\tilde{l}_3}_0 \\ \nonumber
 \end{eqnarray}
 \begin{eqnarray} \label{eq:hartree_diagrams2}
 {\cal D}^{L}_5 &=& (-)^2 \sum_{\substack{ \tilde{l}_1,...,\tilde{l}_5  \\ \sum_i \tilde{l}_i =L} } T\sum_{i\omega}e^{-i\omega0^{-}} \\  \nonumber
 && \;\;\;\;\;\;\; \times 
 \sum_\mathbf{k}  G^{l=1+\tilde{l}_1}_{0,\mathbf{k}}(i\omega) \Sigma^{\tilde{l}_2,\tilde{l}_3,\tilde{l}_4}_{2,\mathbf{k}}(i\omega) G^{l=1+\tilde{l}_5}_{0,\mathbf{k}}(i\omega)
\end{eqnarray}

As we are restricting to $K\leq 5$ calculations, the ${\cal D}^{L}_{3...5} $ insertions can only be added once, and only with $L=0$. We now define $M^{L}_i$ as the number of insertions of ${\cal D}^{L}_i$ tadpoles, and we define $N_i$ as the number of interaction vertices contained in the tadpole ${\cal D}_i$ (regardless of $L$, we have $N_1=1$, $N_2=2$, $N_3=N_4=N_5=3$).

The series can be now resummed as:
\begin{widetext}
\begin{eqnarray} \label{eq:bare_resummation}
 \Sigma^{(\mathrm{HF})}_{\mathbf{k}}(z) 
 &=& \sum_{K=2}^{\infty} \sum_{N=2}^{K} \sum_{L=0}^{K-N} \sum^{K-N-L}_{\substack { \{M^{L'}_i\} =0 \\ 
                                         N+L+\sum_{i,L'} M^{L'}_i(N_i + L') = K
                                       }} \times \\ \nonumber
&&\;\;\;\;\;\;\;  \times              (-U)^{N+\sum_{i,L'} M^{L'}_i N_i} (-\delta\mu)^{L+\sum_{i,L'} M^{L'}_i L'} \prod_{i,L'} \left( {\cal D}^{L'}_i\right)^{M^{L'}_i}
                                       \Omega(L,\{M_i^{L'}\}) \sum_{\Upsilon_N}  D_{\Upsilon_N,\mathbf{k},L+\sum_{i,L'} M^{L'}_i}(z) 
\end{eqnarray}
\end{widetext}
where $\Omega(L,\{M_i^{L'}\})$ is the combinatorial prefactor which counts all possible ways the selected single-particle vertices $\delta\mu$,$\{{\cal D}_i\}$ can be arranged. This corresponds to the number of permutations of multisets
\begin{equation}
\Omega(L,\{M_i^{L'}\}) = \frac{(L+\sum_{i,L'} M^{L'}_i)!}{L!\prod_{i,L'} M^{L'}_i !}
\end{equation}
We emphasize that Eq.~\ref{eq:bare_resummation} is fully general, but at orders $K\geq5$, additional Hartree insertions $\cal D$ (compared to Eqs \ref{eq:hartree_diagrams}-\ref{eq:hartree_diagrams2}) need to be considered.

Finally, we stress that our analytical time-integral solution and action-shift tuning scheme in DiagMC are not restricted to the treatment of the Hubbard Hamiltonian. See Appendix~\ref{sec:general_hamiltonian} for a discussion of DiagMC in the case of a general Hamiltonian with 2-body interactions.

\section {Results} \label{sec:results}

\begin{figure*}[ht!]
 \begin{center}
 \includegraphics[width=6.4in, trim=0 0 0 0, clip]{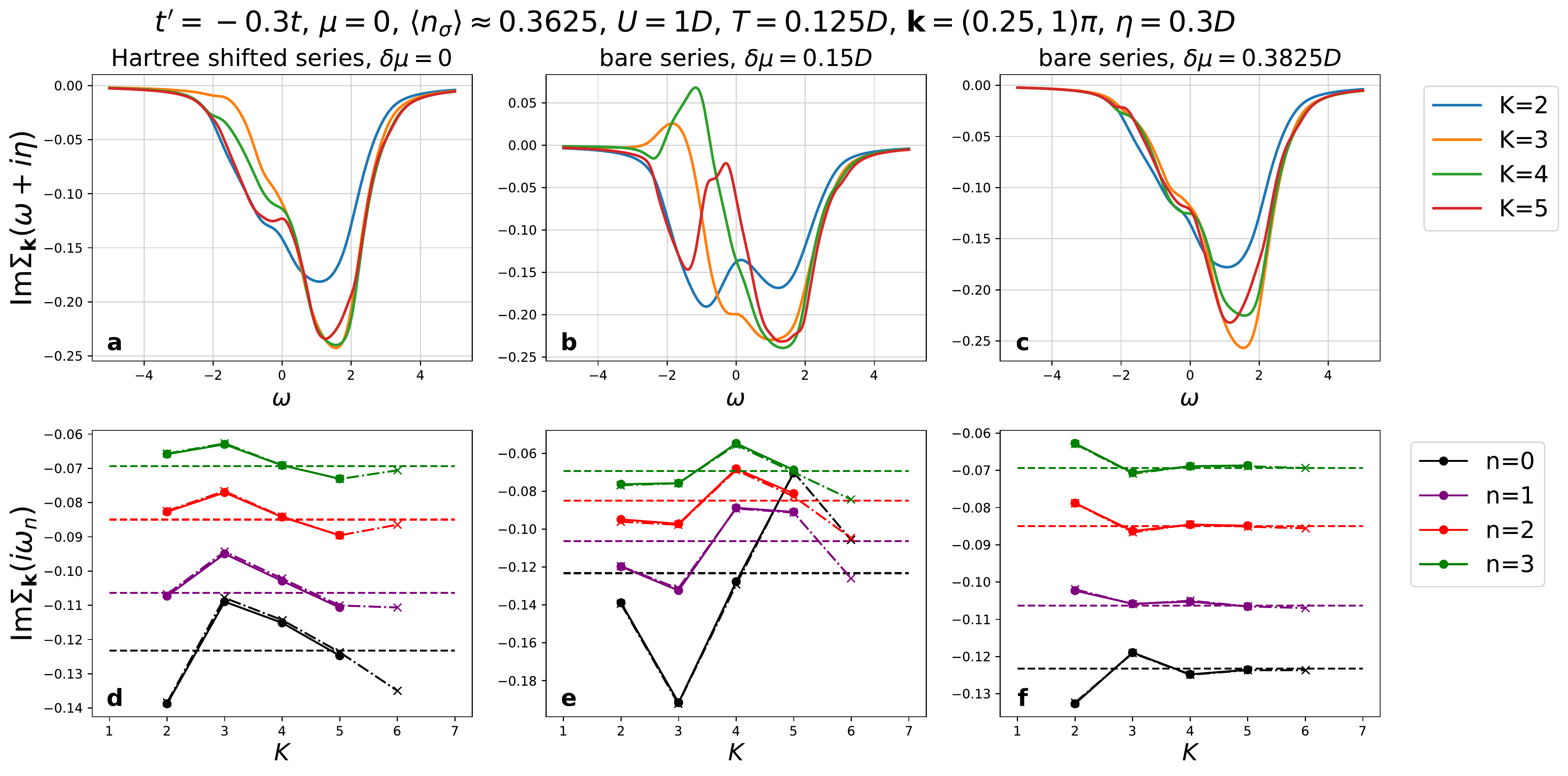}
  \end{center}
 \caption{DiagMC solution for the Hubbard model on a square lattice. Top row: imaginary part of self-energy on the real axis (with broadening $\eta$) obtained with three different series, up to perturbation order $K$. Bottom row: illustration of convergence with respect to perturbation order $K$, using values of the imaginary part of self-energy at the lowest four Matsubara frequencies $i\omega_{n=0..3}$. Full lines is our result, dash-dotted lines with crosses is the analogous result with a numerical $\tau$-integration algorithm from Ref.\onlinecite{wuPRB2017}, and horizontal dashed lines is determinantal QMC result on a $16\times 16$ lattice from Ref.\onlinecite{wuPRB2017}.
 }
 \label{fig:main}
\end{figure*}

\subsection{Convergence speed-up with $\delta\mu$ expansion in the bare series}

Here we focus on supplementing the results from Ref.\onlinecite{wuPRB2017} with
real-frequency self-energies calculated without any numerically ill-defined
analytical continuation.

The model parameters are $t'=-0.3t$, $\mu=0$, $U=1.0D$, $T=0.125D$, and $\langle n_\sigma \rangle=0.3625$. In Ref.\onlinecite{wuPRB2017}, the calculation was performed with the Hartree shifted series with $\delta\mu=0$, as well as with the bare series, with two values of $\delta\mu$, namely $0.15D$ and $0.3825D$. We repeat these calculations with our method. We use lattice size $32\times 32$, and project the dispersion on a uniform energy grid as described in Ref.\onlinecite{VucicevicPRB2020}, and discussed in Section \ref{sec:numerical_evaluation}. In Fig.\ref{fig:main} we show our results, and compare them with the results of Ref.\onlinecite{wuPRB2017}.

In the upper row of Fig.\ref{fig:main} are the real-frequency self-energies calculated up to order $K\leq 5$. We are keeping a finite broadening $\eta=0.3D$, to smoothen the curves. As discussed in Appendix~\ref{sec:broadening}, in our method, numerical pole-broadening is not a formal necessity. However, there is still a significant amount of statistical noise in our real-frequency result (although the imaginary-frequency result is already very well converged). It is important to note that some of the noisy features in our real-frequency result may be artifacts of the finite lattice size that would not vanish with increasing number of MC steps. However, by comparing the result with a $256\times 256$ lattice calculation (Appendix \ref{sec:latt_size}) we check that at already at $\eta=0.2D$, no such artifact should be visible. 
It appears that for the given external $\mathbf{k}$ and broadening $\eta=0.2D$, increasing the lattice size further from $32\times 32$ brings no new information, but it also does not present an additional cost: at $\eta=0.2D$, our $256\times 256$ lattice calculation appears equally well converged as the $32\times 32$ lattice calculation, with the equal number of MC steps and a similar runtime and yields a result that is on top of the $32\times 32$ calculation.
% For a given external momentum $\mathbf{k}$ and a sufficiently big $\eta$ so that finite-size effects are washed out, the lattice size can be increased arbitrarily at no apparent cost: at $\eta=0.2D$, the $256\times 256$ lattice calculation appears equally well converged as the $32\times 32$ lattice calculation with the equal number of MC steps and similar runtime, and yields a result that is on top of the $32\times 32$ calculation.

In the bottom row of Fig.\ref{fig:main} we show the change in the imaginary part of the self-energy at the first 4 Matsubara frequencies, as a function of the maximal order $K$. Full-line and dots is the result of our calculations. The dash-dotted lines with crosses are data points taken from Ref.\onlinecite{wuPRB2017}. The horizontal dashed lines is the $16\times 16$-lattice determinantal QMC result, also from Ref.\onlinecite{wuPRB2017}.

The excellent agreement with the results from Ref.\onlinecite{wuPRB2017} serves as a stringent test of our implementation. In the $\delta\mu=0.3825D$ calculation, even on the real-axis, the self-energy does appear well converged by order $K=5$, although there is some discrepancy between $K=4$ and $K=5$ at around $\omega=1.5D$. 

\begin{figure}[ht!]
 \begin{center}
 \includegraphics[width=2.6in, trim=0.cm 0cm 0cm 0, clip]{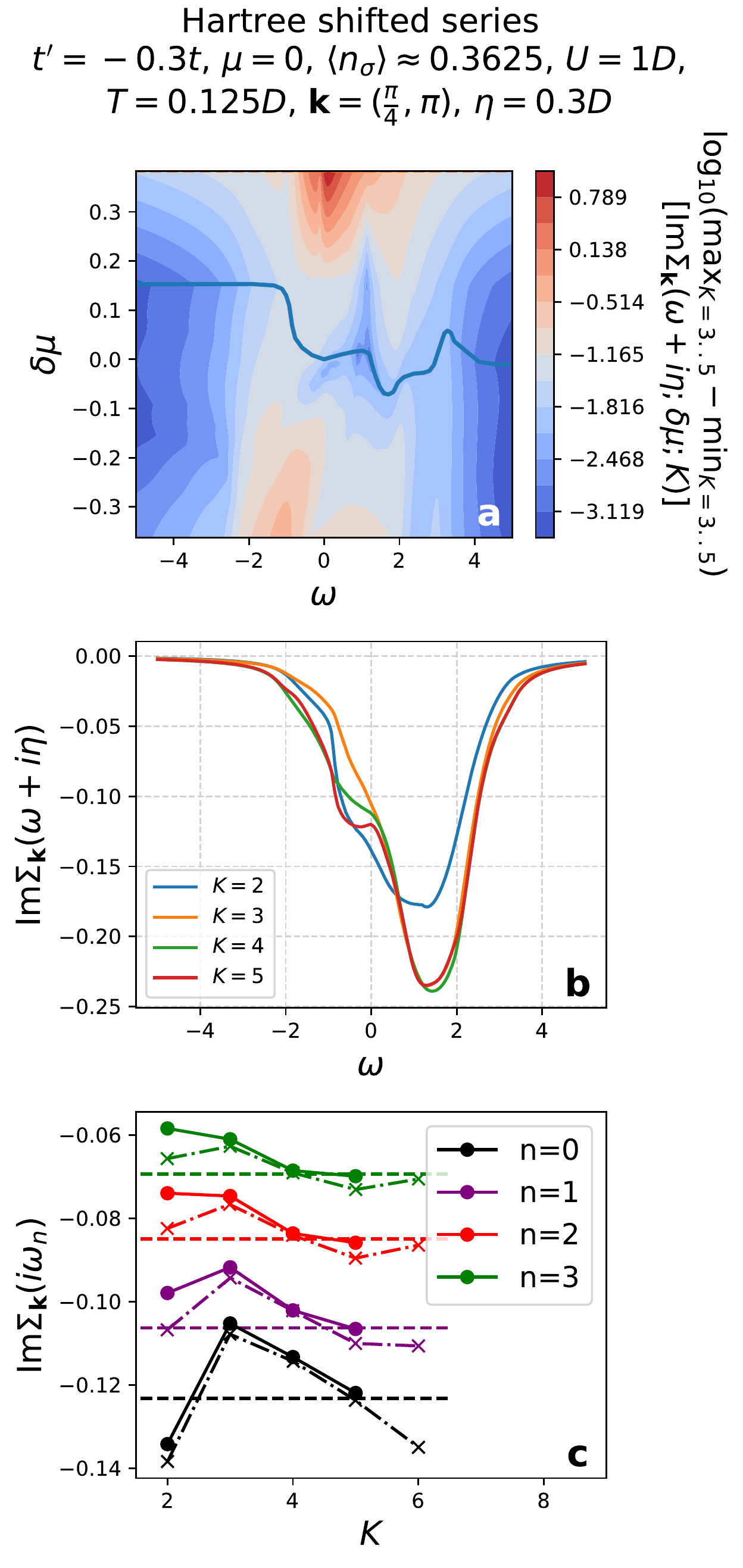}
  \end{center}
 \caption{Results of the Hartree shifted series with $\omega$-resolved resummation, to be compared to Fig.\ref{fig:main}$\mathbf{a}$ and $\mathbf{d}$. Top panel: color plot of the spread of the imaginary part of the self energy at a given $\omega+i\eta$ between orders $K=3$ and $5$, in a calculation with a given $\delta\mu$. Blue line smoothly connects the minima of the spread (at each $\omega$), and defines the $\omega$-dependent optimal shift $\delta\mu^*(\omega)$ used in the resummation. Middle and bottom panel are analogous to Fig.\ref{fig:main}$\mathbf{a}$ and $\mathbf{d}$. In the bottom panel, the dashed-dotted and the dashed lines are the same as in Fig.\ref{fig:main}$\mathbf{d}$.
 }
 \label{fig:hf_series_dmustar}
\end{figure}

\begin{figure}[ht!]
 \begin{center}
 \includegraphics[width=2.6in, trim=0.cm 0cm 0cm 0, clip]{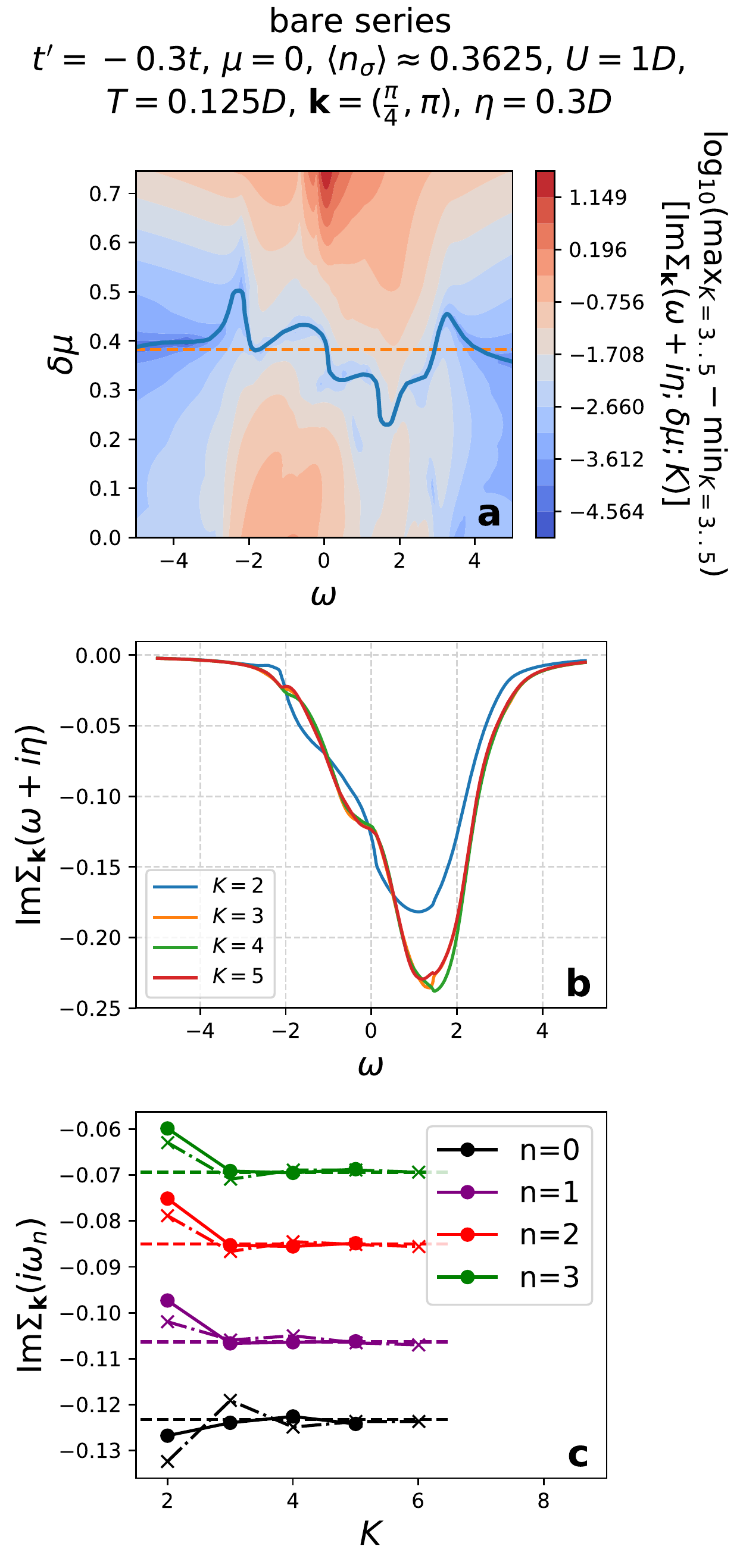}
  \end{center}
 \caption{Results of the bare series with $\omega$-resolved resummation, to be compared to Fig.\ref{fig:main}$\mathbf{c}$ and $\mathbf{f}$. Top panel is analogous to Fig.\ref{fig:hf_series_dmustar}$\mathbf{a}$. Horizontal orange dashed line denotes the value of $\delta\mu$ used in Fig.\ref{fig:main}$\mathbf{c}$ and $\mathbf{f}$ to best converge the imaginary-axis result. Middle and bottom panel are analogous to Fig.\ref{fig:main}$\mathbf{c}$ and $\mathbf{f}$. In the bottom panel, the dashed-dotted and the dashed lines are the same as in Fig.\ref{fig:main}$\mathbf{f}$.
 }
 \label{fig:bare_series_dmustar}
\end{figure}

\begin{figure*}[ht!]
 \begin{center}
 \includegraphics[width=6.4in, trim=0 1cm 0 0, clip]{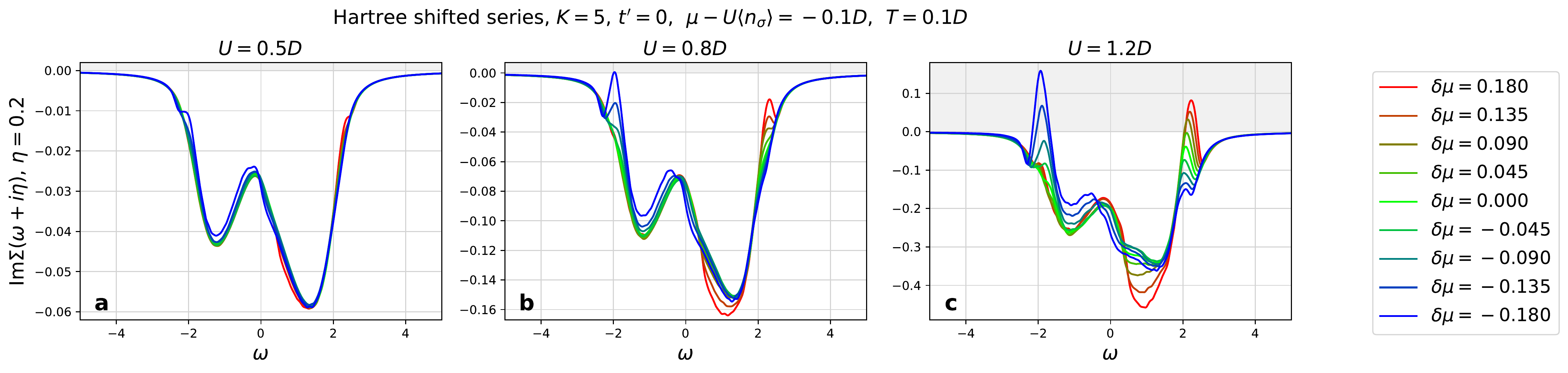}
  \end{center}
 \caption{Imaginary part of the self-energy on the real axis (with broadening $\eta$), at different values of coupling constant $U$, obtained with our method at $K=5$, using different chemical potential shifts $\delta\mu$. The parameters of the calculation are the same as in Ref.\onlinecite{VucicevicPRB2020}. Passing of the curves through the gray shaded area indicates breaking of causality.
 }
 \label{fig:dmus}
\end{figure*}

\begin{figure}[ht!]
 \begin{center}
 \includegraphics[width=2.4in, trim=0cm 0cm 0cm 0, clip]{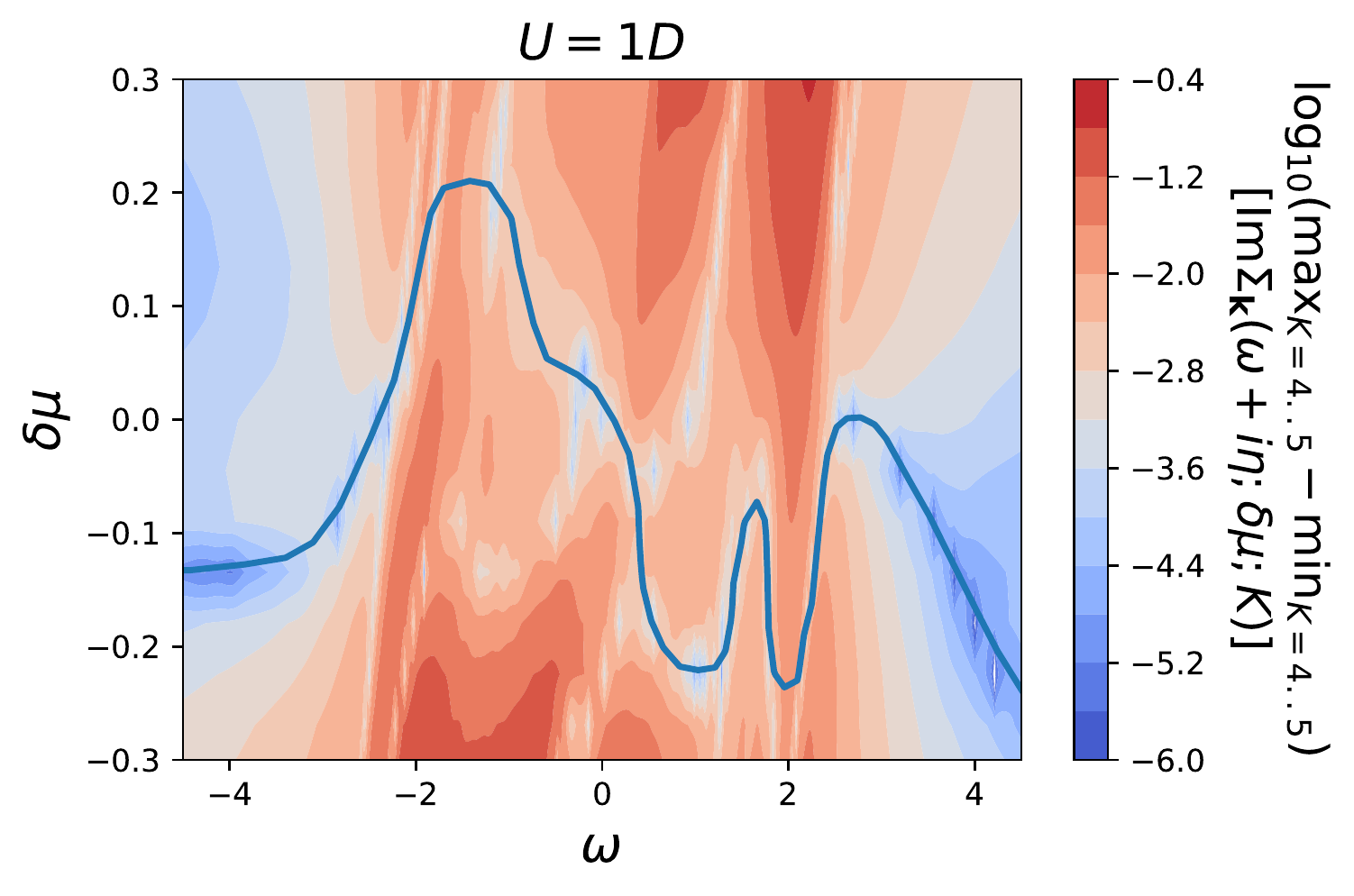}
  \end{center}
 \caption{Analogous to Fig.\ref{fig:hf_series_dmustar}$\mathbf{a}$, for the parameters of the model corresponding to Fig.\ref{fig:dmus}. The blue line is the optimal $\delta\mu^*$, to be used in Fig.\ref{fig:dmu_star_averaging}.
 }
 \label{fig:dmu_star_available}
\end{figure}

\begin{figure}[ht!]

 \begin{center}
 \includegraphics[width=3.2in, trim=1.cm 2.5cm 2.5cm 0, clip]{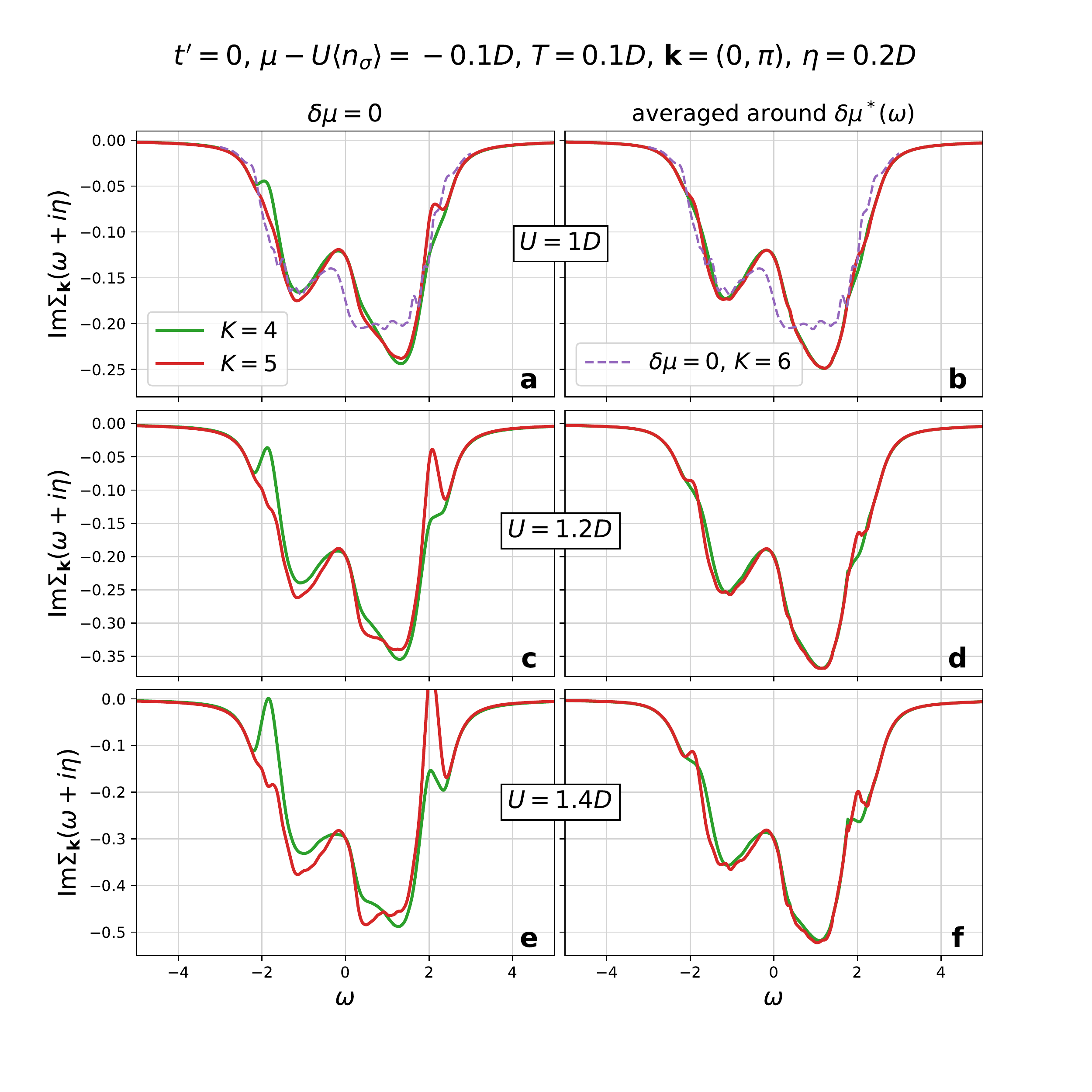}
  \end{center}
 \caption{Imaginary part of self-energy, real-frequency results (with broadening $\eta$). Right column: obtained with the $\omega$-resolved resummation for the model parameters from Fig.\ref{fig:dmus}, using the optimal $\delta\mu^*(\omega)$ from Fig.\ref{fig:dmu_star_available}; To be compared to the standard $\delta\mu=0$ calculation in the left column. Purple dashed lines in top row are the $K=6$ calculation with $\delta\mu=0$.
 }
 \label{fig:dmu_star_averaging}
\end{figure}

\begin{figure}[ht!]
 \begin{center}
 \includegraphics[width=3.2in, trim=0.cm 0cm 0cm 0, clip]{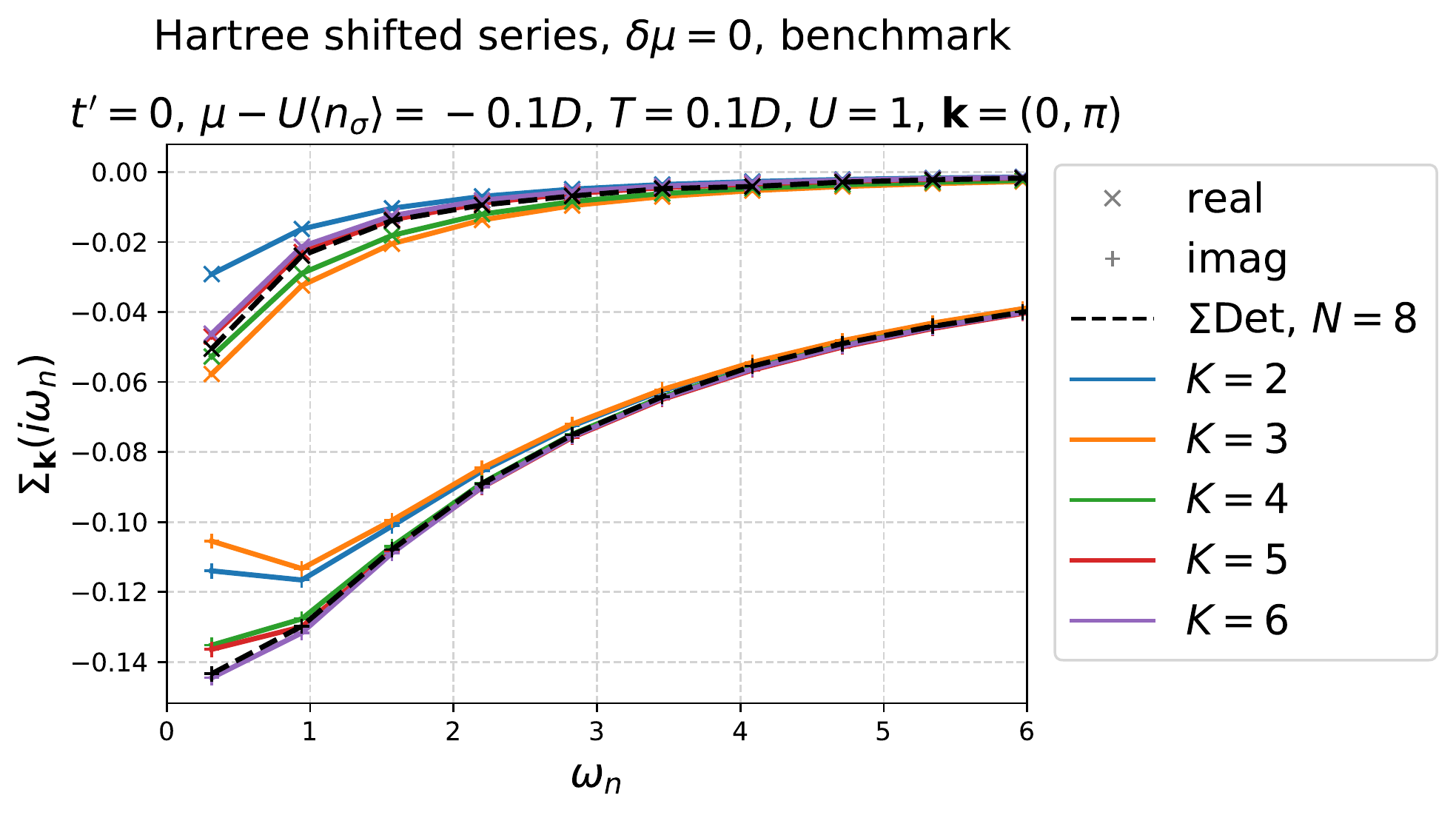}
  \end{center}
 \caption{Matsubara-frequency self-energy result, with model parameters as in Fig.\ref{fig:dmus}. Crosses are the real part, pluses are the imaginary part, lines are eye-guides. Solid lines are the Hartree-shifted series with $\delta\mu=0$ at different maximal $K$. The same result was obtained with both the algorithm presented in this work, as well as with the algorithmic Matsubara summation method from Ref.\onlinecite{VucicevicPRB2020} (two methods were compared diagram by diagram). Black dashed lines is the $\Sigma$Det result at maximal order $N=8$.
 }
 \label{fig:order6_benchmark}
\end{figure}

\subsection{$\omega$-resolved resummation}

We can now go one step further by resumming the series presented in Fig.\ref{fig:main}a and Fig.\ref{fig:main}c for each $\omega$ individually, using an $\omega$-dependent optimal shift $\delta\mu^*(\omega)$.

The results are shown in Figs.\ref{fig:hf_series_dmustar} and \ref{fig:bare_series_dmustar}.

We determine the optimal $\delta\mu^*(\omega)$ by minimizing the spread of the $\mathrm{Im}\Sigma(\omega+i\eta)$ results between orders $K=3$ and $K=5$.
This spread as a function of $\omega$ and $\delta\mu$ is color-plotted in Figs.\ref{fig:hf_series_dmustar} and \ref{fig:bare_series_dmustar}.
We have results for a discrete set of $\delta\mu\in\{\delta\mu_i\}$, so the optimal $\delta\mu^*(\omega)$ is a priori a discontinuous curve. 
As this is clearly non-satisfactory, we smoothen the curve (shown with the blue line on the top panels in Figs.\ref{fig:hf_series_dmustar} and \ref{fig:bare_series_dmustar}).
However, we do not have results for each precise value of this optimal $\delta\mu^*(\omega)$. One could take for each $\omega$ the available $\delta\mu_i$ that is closest to $\delta\mu^*(\omega)$, but this would, again, result in a discontinuous curve. To avoid this, we average the available results as
\begin{equation}
 \Sigma(\omega) = \frac{\sum_i \Delta\delta\mu_i \Sigma(\omega;\delta\mu_i) w(\delta\mu^*(\omega), \delta\mu_i)}{\sum_i \Delta\delta\mu_i w(\delta\mu^*(\omega), \delta\mu_i) }
\end{equation}
where $\Delta\delta\mu_i$ is the size of the $\delta\mu$-step in the available results at the $i$-th value (allows for non-uniform grids).
We use a narrow Gaussian weighting kernel
\begin{equation}
 w(\delta\mu^*(\omega), \delta\mu_i) = e^{-(\delta\mu_i-\delta\mu^*(\omega))^2/W^2}
\end{equation}
The width of the kernel $W$ is chosen such that it is as narrow as possible, 
while still encompassing at least 3-4 $\delta\mu_i$-points, so that the final result is reasonably smooth as a function of $\omega$; $W$ is therefore determined according to the resolution in $\delta\mu$. We use $W=0.05$ and $\Delta\delta\mu_i\approx 0.02$ and have checked that the results are insensitive to the precise choice of this numerical parameter.

The results of the averaging around the optimal $\delta\mu^*(\omega)$ are shown in the middle and bottom panels of Figs.\ref{fig:hf_series_dmustar} and \ref{fig:bare_series_dmustar}. In both cases, the $\omega$-resolved resummation helps to converge the result. In the case of the bare series, the convergence is now almost perfect, and already order $K=3$ is on top of the exact result. In the case of the Hartree shifted series, the results are not perfectly converged at $\omega<0$, yet the $K=5$ calculation is practically on top of the exact result on the imaginary axis, and presents an improvement to the $\delta\mu=0$ series in Fig.\ref{fig:main}a. Note that the improvement in convergence is seen on the imaginary axis, as well.

\subsection{Removing non-physical features}

In this section we focus on the parameters case discussed in Ref.\onlinecite{VucicevicPRB2020}. We calculate the Hartree-shifted series with parameters of the model $t'=0$, $\mu-U\langle n_\sigma\rangle = -0.1D$, $T=0.1D$, and employ various $\delta\mu$ shifts. The lattice size is again $32\times 32$ and we focus on the self-energy at $\mathbf{k}=(0,\pi)$. Note that in Hartree-shifted series, the quantity that enters the calculation is $\mu-U\langle n_\sigma\rangle$, rather than $\mu$. If $\langle n\rangle$ is calculated, $\mu$ can be estimated a posteriori. In our calculation we fix $\mu-U\langle n_\sigma\rangle$, and $\langle n_\sigma\rangle$ is then $U$-dependent. Roughly, as given in Ref.\onlinecite{VucicevicPRB2020}, at $U=1$, we have $\langle n_\sigma \rangle \approx 0.455$.

The results are presented in Fig.\ref{fig:dmus} for three values of $U$. 

At low $U$, the series is well converged by $K=5$, and the result is entirely insensitive to the choice of $\delta\mu$, as expected.

At intermediate and high $U$, the result can be strongly $\delta\mu$ sensitive. The $\delta\mu$ dependence of the result, however, strongly varies with $\omega$. It appears that for a given $\omega$, there are ranges of $\delta\mu$ value where the result (at fixed order $K$) is insensitive to the precise choice of $\delta\mu$. This presents an alternative way of choosing an optimal $\delta\mu$ (similar idea was employed in a different context in Ref.~\onlinecite{AyralPRL2017}).

The striking feature at large $U$ is the causality violations at $|\omega|\approx 2$ that were previously discussed in Ref.\onlinecite{VucicevicPRB2020} (note that the broadening somewhat masks the extent of the problem). The dips in the self-energy spectrum appear to happen only at certain values of $\delta\mu$: at $\omega=-2$, the problem is present at $\delta\mu$ large and negative, and at $\omega=2$ at $\delta\mu$ large and positive. In particular at $\omega=2$, the result appears to vary uniformly with $\delta\mu$, and one cannot select an optimal $\delta\mu$ based on sensitivity of the result to the $\delta\mu$ value. We therefore repeat the procedure from the previous section and select the optimal $\delta\mu^*(\omega)$ based on level of convergence between orders $K=4$ and $K=5$. The spread of results and a smooth choice of $\delta\mu^*(\omega)$ are presented in Fig.\ref{fig:dmu_star_available}.

In Fig.\ref{fig:dmu_star_averaging}, the results of the averaging are shown and
compared to $\delta\mu=0$ results at the highest available orders $K=4$ and
$K=5$, at three values of $U$.  The convergence is visibly better around our
$\delta\mu^*$ than with $\delta\mu=0$ at problematic frequencies
$|\omega|\approx 2$. More importantly, the non-physical features are clearly
absent. At $U=1$, in the $\delta\mu=0$ calculation, the causality is not yet
violated, but the dip at $\omega=2$ is already starting to appear, which is
clearly an artifact of the series truncation which should be removed
systematically. It is important that the intermediate frequency behavior we
obtained by averaging results around the optimal $\delta\mu$ is indeed the
correct one, and it will not change much further with increasing orders. We
show in the top panels the $K=6$, the $\delta\mu=0$ result which has been
benchmarked against a fully converged imaginary-axis result in
Fig.\ref{fig:order6_benchmark} (the converged result was obtained with the
$\Sigma$Det method\cite{moutenet2018,fedor_2017} at order 8). Clearly, the improved
convergence between orders 4 and 5 that we have achieved by choosing
$\delta\mu$ appropriately, does indeed mean an improved final result. However,
our procedure does not improve the result at around $\omega=0$ where the
optimal $\delta\mu$ does appear to be close to 0. The $K=6$, $\delta\mu=0$
result shown in the upper panels of Fig.\ref{fig:dmu_star_averaging} is still a
bit different from $K=5$, $\delta\mu\approx\delta\mu^*(\omega)$ results around
$\omega=0$. 

In the case of $U=1D$, it is interesting that a large negative $\delta\mu$ does bring the $\omega\approx 0$ result at order $K=5$ much closer to the exact value. This can be anticipated from Fig.\ref{fig:dmus} where we show corresponding results for $U=0.8D$ and $U=1.2D$.
Also, by looking at the color plot in Fig.\ref{fig:dmu_star_available}, we see that at $\omega=0$, there is indeed a local minimum in the spread at around $\delta\mu=-0.2$ which could be used as the optimal $\delta\mu^*$. This minimum, however, cannot be continuously connected with the other minima that we observe at $\omega<0$, so we chose a different trajectory in the $(\omega,\delta\mu)$-space. It would be interesting for future work to inspect the behavior at even more negative $\delta\mu$, where another continuous trajectory $\delta\mu^*(\omega)$ might be found.

\section{Discussion, conclusions and prospects} \label{sec:conclusion}

In this paper we have derived an analytical solution for the multiple-time
integral that appears in the imaginary-time Feynman diagrams of an interaction
series expansion.  The solution is general for any diagram with a single
external time, or no external times. We find this generality to be a great
advantage compared to the recently proposed algorithmic solutions of the
corresponding Matsubara-frequency summations. Our analytical solution allowed
us to develop a very flexible DiagMC algorithm that can make use of the
possibility to optimize the series with shifted actions. As a result we were
able to almost perfectly converge a real-frequency self-energy in just 3-4
orders of perturbation, in a non-trivial regime and practically in the thermodynamic limit.

More importantly, the fact that one does not have to prepare a solution for
each diagram topology individually opens the possibility to develop algorithms
more akin to CTINT and allow the MC sampling to go to indefinite perturbation orders. In
fact, upon a simple inspection of CTINT and segment-CTHYB
equations\cite{gullRMP2011}, it becomes clear that our solution can in
principle be applied there, so as to reformulate these methods in
real-frequency. This would, however, come at the price of having to break into individual terms the determinant 
that captures all the contributions to the partition function at a given perturbation order.
In turn, this may lead to a more significant sign problem, and an effective cap on the perturbation orders
that can be handled in practice. On the other hand, it is not entirely clear 
how much of the sign problem comes from summing the individual terms, and how much from
the integration of the internal times, and we leave such considerations for future work.
In any case, DiagMC algorithms based on hybridization expansion have been proposed before (see Ref.\onlinecite{GullPRB2010,cohenPRL2015,edelstein2019}),
where our analytical solution may be applied.

Our solution also trivially generalizes to real-time integrals and may have use
in Keldysh and Kadanoff-Baym\cite{AokiRMP2014} calculations, where the infamous
dynamical sign problem arises precisely due to oscillating time integrands.
There have been recent works\cite{KunitsaPRE2020,HutcheonPRE2020} with
imaginary-time propagation of randomized walkers where our solution may also find
application.

Finally, we emphasize that avoiding analytical continuation 
%becomes particularly important
could be beneficial
at high temperature where the Matsubara frequencies become distant from the real axis, 
and thus noisy imaginary-axis correlators contain little information.\cite{VucicevicPRL2019,HuangScience2019}
The high-temperature regime is particularly relevant for optical lattice simulations of the Hubbard model.\cite{BrownScience2018}
In that context, we anticipate our method will find application in the calculation of conductivity
and other response functions.

\appendix

\section{Real-time integration} \label{sec:real_time}

Let's consider the following special case of the integral Eq.\ref{eq:integral}, which is relevant for real-time integrations featuring integrands of the form $e^{itE}$:
\begin{equation}\label{eq:real_time_integral}
 \tilde{{\cal I}}_{\{l_2...l_N\}, \{E_2...E_N\}}(t) = \prod_{i=2}^{N}
 \int_{0}^{t_{i+1}} \mathrm{d}t_i\; t^{l_i}_i \;
  e^{it_{i}E_i} 
\end{equation}
with $t_{N+1}\equiv t$. This corresponds to the case $r\notin[2,N]$ in Eq.\ref{eq:integral}, and $\omega_i = iE_i $, and we will define $\tilde{E}_i$ analogously to $\tilde{\omega}_i$. The result is then obtained straightforwardly from Eq.\ref{eq:main_analytical_solution}
\begin{eqnarray} \label{eq:real_time_analytical_solution}
&&  \tilde{{\cal I}}_{\{l_2...l_N\}, \{E_2...E_N\}}(t) = \\ \nonumber
&& \;\;\;      \sum_{\{b_i\in[\delta_{\tilde{z}_i},1]\}_{i=2..N}}
   e^{i t \tilde{E}_{N} b_N }
  \sum_{\{k_i\in[0,(1-\delta_{\tilde{z}_i})n_i]\}_{i:b_i=1}}   
   \\ \nonumber
&& \;\;\;\times      
(-1)^{\sum_{i=2}^{N} k_i} \prod_{i: \delta_{\tilde{z}_{i}}=1}\;\; \frac{1}{n_{i}} \\ \nonumber
&& \;\;\;\times  t^{n_{N}+1-b_{N}-k_{N}}
  \prod_{
    i: \tilde{E}_{i}\neq0
  }\frac{C_{n_i,k_i}}{(i\tilde{E}_i)^{k_i+b_i}} \\ \nonumber
\end{eqnarray}
which has the following general form
\begin{equation}
 \tilde{{\cal I}}(t) = \sum_{j;p\in\mathbb{N}_0} {\cal Z}_{p,j} t^p e^{i t {\cal E}_j} \;\;.
\end{equation}

\begin{figure}[ht!]
 \begin{center}
 \includegraphics[width=2.9in, trim=0.cm 0cm 0cm 0, clip]{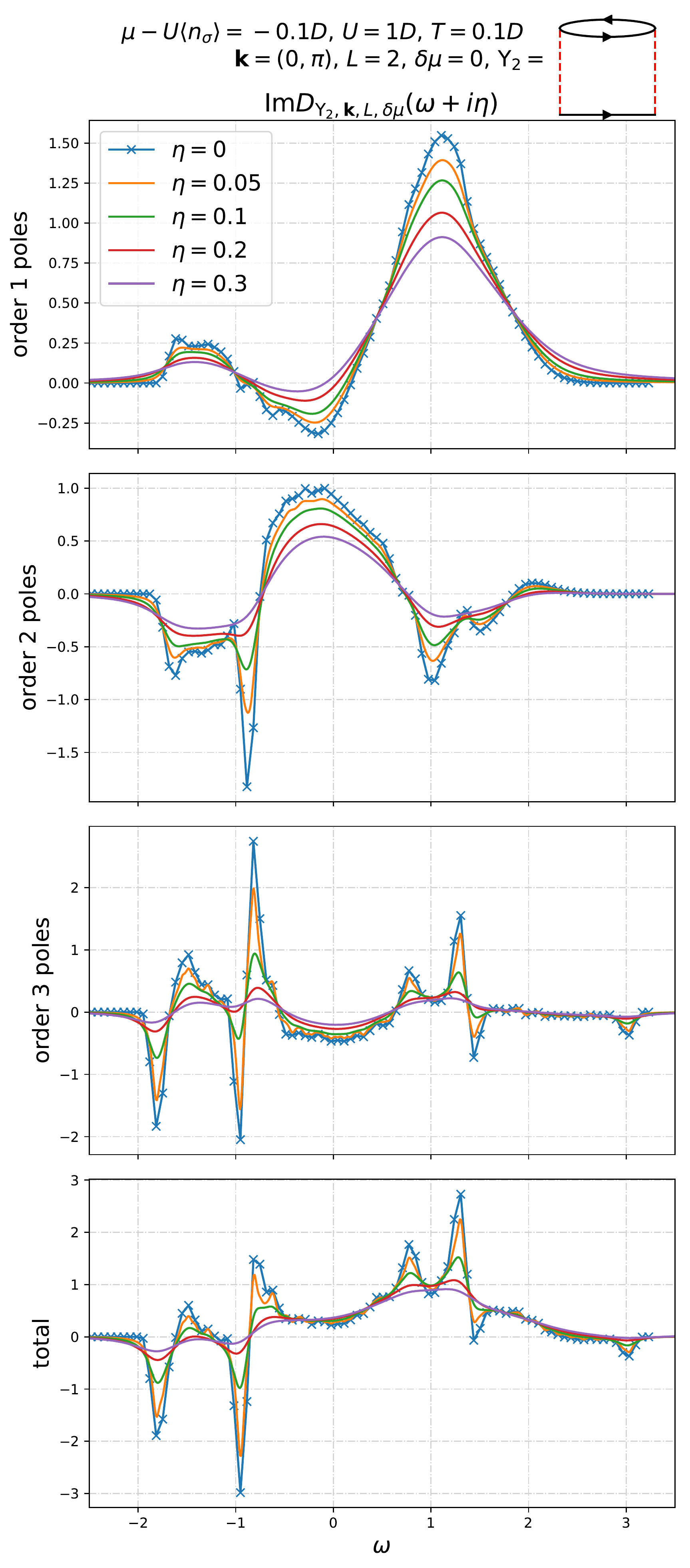}
  \end{center}
 \caption{Illustration of an $\eta=0^+$ result obtained without any numerical broadening, based only on pole amplitudes. Diagram used is the second order diagram, with $L=2$ (illustrated at the top). Top three panels are contributions from 1st, 2nd and 3rd order poles, respectively. Bottom panel is the total result. Lines with $\eta>0$ are obtained with numerical broadening. The crosses on the $\eta=0$ result denote the available frequencies (in between we assume linear interpolation).
 }
 \label{fig:eta0}
\end{figure}

\begin{figure}[ht!]
 \begin{center}
 \includegraphics[width=2.9in, trim=0.cm 0cm 0cm 0, clip]{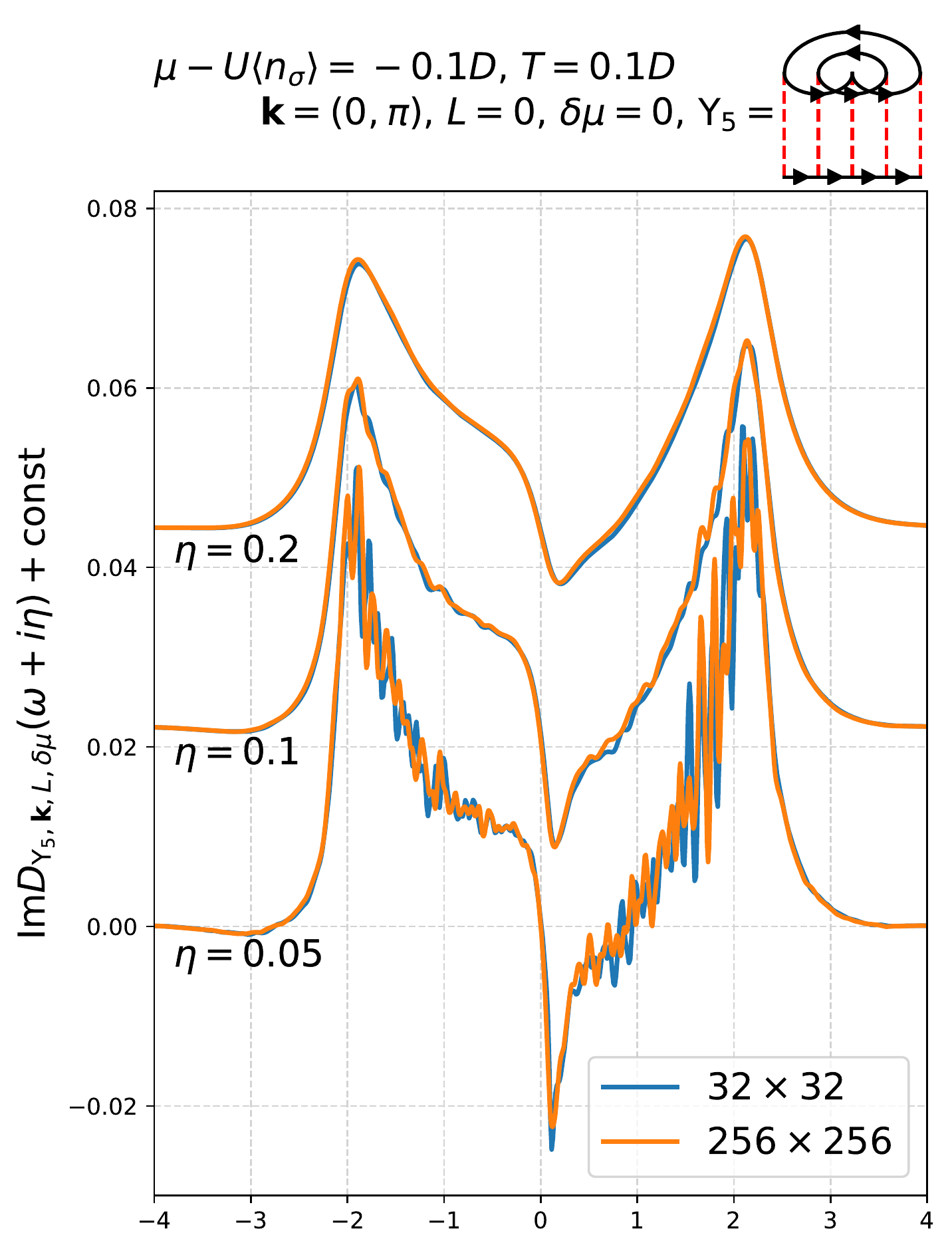}
  \end{center}
 \caption{Comparison of the real-frequency imaginary self-energy result for a single 5th order diagram, for the lattice sizes $32\times 32$ and $256\times 256$, at three different levels of broadening. The calculation is in both cases performed with the same number of MC steps.
 }
 \label{fig:latt_size}
\end{figure}

\section{Extracting real-axis results without pole-broadening}\label{sec:broadening}

In this section we show how the results on the real-axis can be extracted
without any numerical broadening of the poles. Rather, we make use of the pole
amplitudes, by interpreting the result as being representative of the
thermodynamic limit, where poles on the real-axis merge into a branch cut, thus
we consider that the pole amplitude is a continuous function of the
real-frequency. We extract the imaginary part of the contribution ($\mathrm{Im}D(\omega)$), and then the
Hilbert transform can be used to reconstruct the real part.

The procedure relies on the following construction. A function $f(z)$ which is analytic everywhere in the upper half of the complex plane ($z^+=x+iy$ with $y>0$) and decays to zero with $|z^+|$ satisfies the relation
\begin{equation}
 f(z^+) = -\frac{1}{\pi}\int \mathrm{d}x' \frac{\mathrm{Im}f(x'+i0^+)}{z^+-x'}
\end{equation}
After applying the $p$-th derivative with respect to $x$ (i.e. the real part of $z^+$) on both sides of the equation, one obtains:
\begin{eqnarray}
 \partial^p_x f(z^+) &=& -\frac{1}{\pi}\int \mathrm{d}x' \partial^p_x \frac{\mathrm{Im}f(x'+i0^+)}{z^+-x'} \\ \nonumber
                     &=& -\frac{1}{\pi}\int \mathrm{d}x' (-1)^p (p+1)! \frac{\mathrm{Im}f(x'+i0^+)}{(z^+-x')^{p+1}}
\end{eqnarray}
We can now move the constant prefactors to the left-hand side and rename $p+1\rightarrow p$. Just above the real axis we have
\begin{eqnarray} \nonumber
 \frac{(-1)^{p}\pi}{p!}\partial^{p-1}_x f(x+i0^+)  &=& \int \mathrm{d}x'  \frac{\mathrm{Im}f(x'+i0^+)}{(x-x'+i0^+)^p} \\ 
\end{eqnarray}
We can now discretize the expression on a uniform $x$-grid with the step $\Delta x$, say $x_j = j\Delta x$, and we see that the right-hand side has the form of a sum of poles of order $p$, equidistant along the real-axis, and with amplitudes $ {\cal A}_{j} = \mathrm{Im}f(x_j+i0^+)$
\begin{eqnarray} \nonumber
 \frac{(-1)^{p}\pi}{p!}\tilde{\partial}^{p-1}_j {\cal A}_{j}  &\approx& \mathrm{Im} \sum_{j'}  \Delta x \frac{  {\cal A}_{j'}}{(x_j-x_{j'}+i0^+)^p}, \\ \label{eq:pole_amplitude_use}
\end{eqnarray}
where $\tilde{\partial}$ is the finite-difference approximation for the derivative along the $x$ axis.
Clearly, the imaginary part of the entire sum of $p$-order poles at a certain point $x_j$ can be estimated by looking only at the $p-1$-th derivative of the amplitudes of these poles at $x_j$, as given in the above expression.

The expression Eq.~\ref{eq:pole_amplitude_use} can be readily applied in our case (Eq.\ref{eq:sum_of_poles}) where the real axis is the frequency axis $\omega$, with step $\Delta\omega$ and $\omega_j=j\Delta\omega$, and the sum of poles determines our diagram contribution $D$. In general we have poles of various orders, but we can group the poles by order and treat their contributions separately.
We therefore have
\begin{equation}
\mathrm{Im}D(\omega_j+i0^+) \approx \frac{\pi}{\Delta\omega}\sum_p \frac{(-1)^{p}}{p!}  \tilde{\partial}^{p-1}_j {\cal A}_{j,p}
\end{equation}
In case of simple poles only, the contribution at any $\omega_j$ is simply proportional to the amplitude of the pole ${\cal A}_{j,1}$.
Otherwise, the procedure requires that the pole amplitudes form a reasonably smooth function of real-frequency.
Additionally, the energy resolution is a measure of the systematic error made in this procedure.

\begin{figure*}[]
 \begin{center}
 \includegraphics[width=6.4in, trim=0.cm 0cm 0cm 0, clip]{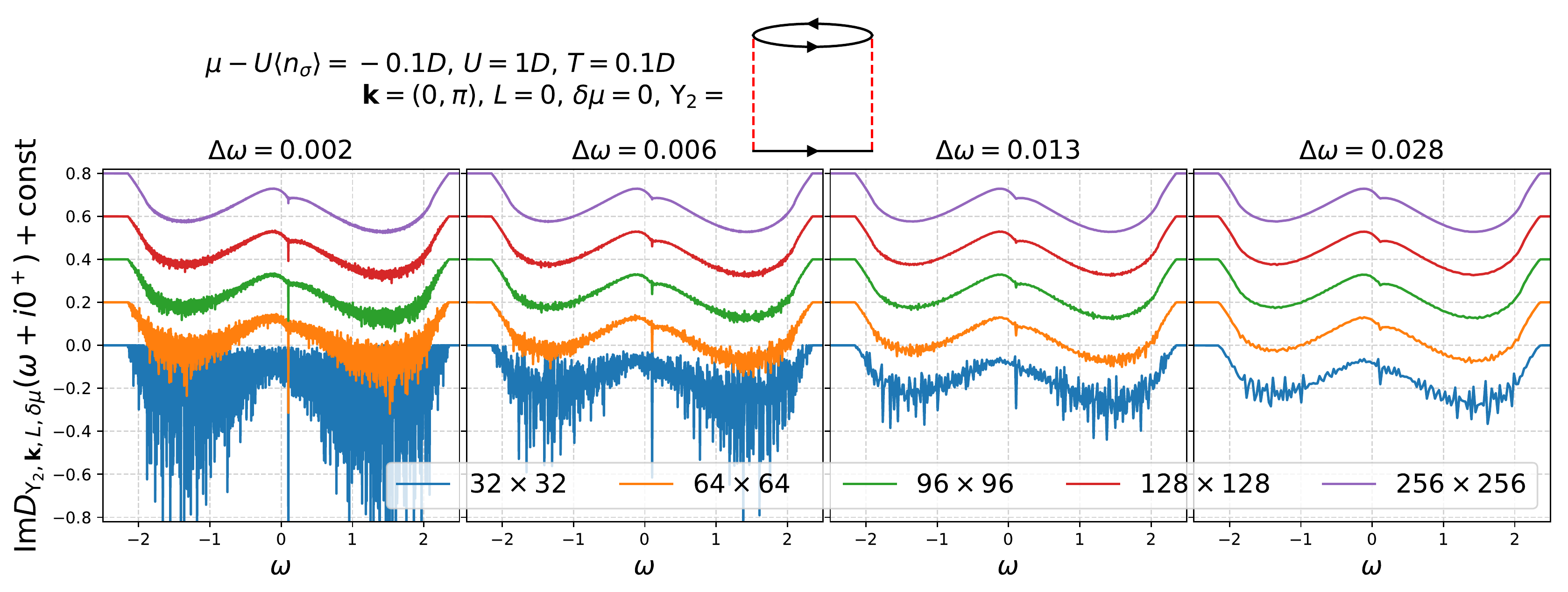}
  \end{center}
 \caption{Real-frequency result ($\eta=0^+$) for the contribution of the lowest order diagram at various lattice sizes and frequency resolutions, obtained with full summation (Gray code). The step of the uniform energy grid is denoted $\Delta\omega$.
 }
 \label{fig:size_and_resolution}
\end{figure*}

To avoid statistical noise and noisy features coming from the finite size of the lattice (see next section), we test our method on the example of a $N=2$, $L=2$ diagram, which we can solve with full summation of Eq.\ref{eq:final_expression}, on a lattice of the size $96\times 96$. This diagram produces poles up to order 3. The result is shown in Fig.\ref{fig:eta0}. In the first three panels we show contribution from poles of each order, and in the bottom panel we show the total result.

\section{Convergence with lattice size}\label{sec:latt_size}

In this section we discuss the convergence of the result with respect to the lattice size.

In Fig.\ref{fig:latt_size} we compare the results for a single $N=5$, $L=0$ diagram on the lattices of size $32\times 32$ and $256\times 256$. We observe that the result is almost exactly the same at broadening level $\eta=0.2$, which brings further confidence in the results in the main part of the paper.

In Fig.\ref{fig:size_and_resolution} we illustrate how the size of the lattice determines the highest energy-resolution one can have, under requirement that the results form a continuous curve on the real axis and are, therefore, representative of the thermodynamic limit. We perform the full summation for the second order diagram with $L=0$, with various sizes of the lattice and various resolutions. Clearly, the bigger the lattice, the higher the energy-resolution one can set without affecting the smoothness of the results.

The numerical parameters of the calculation are therefore the size of the lattice, the energy resolution and the broadening (resolution and the broadening can be tuned a posteriori), and one can tune them to get the optimal ratio between performance and the error bar. If the pole amplitudes ${\cal A}_{jp}$ are a relatively smooth function of $j$, no broadening is then needed at all.

\section{Derivation of Eq.\ref{eq:int1}} \label{sec:proof_eqint1}

After applying $n$ times the partial integration over integral from the left-hand side of Eq.\ref{eq:int1}
we get
\begin{widetext}
\begin{eqnarray}
    \nonumber
    &&\int_0^{\tau_f}\tau^n e^{\tau z}d\tau \\
    \nonumber
    &&\;\;=\frac{1}{z^{n+1}}\int_0^{z\tau_f}\tau^{n}e^{\tau}d\tau\\
    \nonumber
    &&\;\;=\frac{1}{z^{n+1}}\bigg(e^{z\tau_f}(z\tau_f)^n-ne^{z\tau_f}(z\tau_f)^{n-1}+...+(-1)^n n!\int_0^{z\tau_f}\tau^{0}e^{\tau}d\tau\bigg)\\
    \nonumber
   &&\;\;=\frac{1}{z^{n+1}}\left(\frac{n!}{(n-0)!}(-1)^0 e^{z\tau_f}(z\tau_f)^{n-0}+(-1)^{1}\frac{n!}{(n-1)!}e^{z\tau_f}(z\tau_f)^{n-1}+...+(-1)^n\frac{n!}{(n-n)!}\int_0^{z\tau_f}\tau^{0}e^{\tau}d\tau\right)\\
   \nonumber
   &&\;\;=\frac{1}{z^{n+1}}\left(\frac{n!}{(n-0)!}(-1)^0 e^{z\tau_f}(z\tau_f)^{n-0}+(-1)^{1}\frac{n!}{(n-1)!}e^{z\tau_f}(z\tau_f)^{n-1}+...+(-1)^n\frac{n!}{(n-n)!}(z\tau_f)^0(e^{z\tau_f}-1)\right)\\
   \nonumber
   &&\;\;=\frac{1}{z^{n+1}}e^{z\tau_f}\sum_{k=0}^n(-1)^k(z\tau_f)^{n-k}\frac{n!}{(n-k)!}-(-1)^n\frac{n!}{z^{n+1}}\\
\end{eqnarray}
\end{widetext}
which can be readily identified with the right-hand side of Eq.\ref{eq:int1}.

%\clearpage 

\section{Derivation of Eq.\ref{eq:g0_tau}}
\label{sec:g0tau}

We are looking for a solution of the Fourier transform
\begin{equation}
  G_0^l(\varepsilon, \tau) = \frac{1}{\beta} \sum_{i\Omega_\eta}
    \frac{e^{-i\Omega_\eta \tau}}{(i\Omega_\eta - \varepsilon)^l}
\end{equation}
For any $\tau$ we can express the sum above
as a contour integral, and we find
\begin{eqnarray}
    G_0^l(\varepsilon,\tau)
    %&=&\frac{1}{\beta}\sum_{i\Omega_\eta}\frac{e^{-i\Omega_\eta\tau}}{\left(i\Omega_\eta-\varepsilon\right)^l} \\ \nonumber
    &=&-\mathrm{Res}_{z=\varepsilon} \frac{e^{-z\tau}}{\left(z-\varepsilon\right)^l}\frac{\eta^{\left\lfloor\frac{\tau}{\beta}\right\rfloor}
    e^{\left\lfloor\frac{\tau}{\beta}\right\rfloor\beta z}}{1-\eta e^{-\beta z}}dz \\ \nonumber
    &=&  -\frac{\eta^{\left\lfloor\frac{\tau}{\beta}\right\rfloor}}{(l-1)!}
    \left.\frac{d^{l-1}}{dz^{l-1}}\frac{e^{-\beta z\{\frac{\tau}{\beta}\}}}{1-\eta e^{-\beta z}} \right|_{z=\varepsilon}
\end{eqnarray}
where $\lfloor ... \rfloor$ denotes the integer part (floor function), and $\{x\}\equiv x-\lfloor x\rfloor$ denotes the fractional part.

We see that it will be useful to have an expression for
derivatives of $(1-\eta e^z)^{-1}$. They have the
general form
\begin{equation}
  \frac{d^k}{d z^k} \frac{1}{1 - \eta e^z} =
  \sum_{n=0}^k C^k_n \frac{(e^z)^n}{(1 - \eta e^z)^{n+1}}
\end{equation}
By deriving this expression on both sides, one obtains
a recursion for the coefficients $C^k_n$
\begin{equation}
  C^{k+1}_n = n C^k_n + \eta n C^k_{n-1}
\end{equation}
with holds for $k> -1$ and $n> 0$ with $C^{0}_0=1$.
That can be rewritten
\begin{equation}
  \frac{\eta^n}{n!} C^{k+1}_n = n \frac{\eta^n}{n!} C^k_n + \frac{\eta^{n-1}}{(n-1)!} C^k_{n-1}
\end{equation}
If we define $S^k_n = \frac{\eta^n}{n!} C^k_n$ we have the recursion $S^{k+1}_n
= n S^k_n + S^k_{n-1}$, which is the recursion for the Stirling numbers of the
second kind. This allows to have the following important result
\begin{align}
  \nonumber
  \frac{d^k}{d z^k} \frac{1}{1 - \eta e^z} &=
  \sum_{n=0}^k
\eta^n n!
\begin{Bmatrix}
k \\ n
\end{Bmatrix}
\frac{(e^z)^n}{(1 - \eta e^z)^{n+1}} \\
&=
  \sum_{n=0}^k
\eta^n n!
\begin{Bmatrix}
k \\ n
\end{Bmatrix}
\frac{e^{-z}}{(e^{-z} - \eta)^{n+1}}
\end{align}
With this, one obtains the following expression
\begin{eqnarray}\label{eq:final_ft_g0}
G_0^l(\varepsilon,\tau)&=&-e^{\varepsilon\beta\left(1-\left\{\frac{\tau}{\beta}\right\}\right)}\eta^{\left\lfloor\frac{\tau}{\beta}\right\rfloor+1}\left(-\beta\right)^{l-1} 
   \\ \nonumber
&&\;\;\;\times \sum_{m=0}^{l-1}\sum_{n=0}^{l-m-1}\frac{n!}{(l-m-1)!m!}{l-m-1\brace n} \\ \nonumber
&&\;\;\;\;\;\;\;\;\times
\left(\frac{1}{\eta e^{\varepsilon\beta}-1}\right)^{n+1}\left\{\frac{\tau}{\beta}\right\}^{m} \\ \nonumber
\end{eqnarray}

which already satisfies the (anti-)periodicity properties of the Green's function.

To make use of the result Eq.\ref{eq:final_ft_g0}, we need to express
$G_0^l(\varepsilon,\tau)$ as a function of two times
$G_0^l(\varepsilon,\tau,\tau')\equiv G_0^l(\varepsilon,\tau-\tau')$, with
$\tau,\tau' \in [0,\beta]$.
We first consider $\tau\geq\tau'$. By substituting
$(\tau-\tau')^m=\sum_{\zeta=0}^m(-1)^{m-\zeta}\binom{m}{\zeta}\tau^{\zeta}\tau'^{m-\zeta}$
into Eq.\ref{eq:final_ft_g0} and substitute $m-\zeta$ with $\varsigma$ we get
\begin{equation}
    G_0^l(\varepsilon,\tau-\tau')=\eta e^{\varepsilon(\tau'-\tau)}n_{\eta}(-\varepsilon)\sum_{\zeta=0}^{l-1}\sum_{\varsigma=0}^{l-\zeta-1}c_{l,\zeta,\varsigma}^{-}(\varepsilon)\tau^{\zeta}\tau'^{\varsigma}
\end{equation}
with $c_{l,\zeta,\varsigma}^{-}(\varepsilon)$ as defined in
Eq.\ref{eq:coefficients}.
The result for $\tau<\tau'$ can then easily be obtained by proving the
property $G_0^l(\varepsilon,\tau)=(-1)^lG_0^l(-\varepsilon,-\tau)$
\begin{eqnarray}
\nonumber
    G_0^l(\varepsilon,-\tau)
    &=&\frac{1}{\beta}\sum_{n=-\infty}^{\infty}\frac{e^{i\Omega_\eta\tau}}{\left(i\Omega_\eta-\varepsilon\right)^l}\\ \nonumber
    &=&\frac{1}{\beta}\sum_{n=-\infty}^{\infty}\frac{e^{-i\Omega_\eta\tau}}{\left(-i\Omega_\eta-\varepsilon\right)^l}\\  \nonumber
    &=&(-1)^l\frac{1}{\beta}\sum_{n=-\infty}^{\infty}\frac{e^{-i\Omega_\eta\tau}}{\left(i\Omega_\eta+\varepsilon\right)^l} \\ \nonumber
    &=&(-1)^lG_0^l(-\varepsilon,\tau)
\end{eqnarray}
which implies that in the definition Eq.\ref{eq:g0_tau}, we must have
\begin{equation}
  c_{l,\zeta,\varsigma}^{+}(\varepsilon)=(-1)^{l-1}c_{l,\varsigma,\zeta}^{-}(-\varepsilon).
\end{equation}

\section{General Hamiltonian case} \label{sec:general_hamiltonian}

It is important to show that our method is not restricted to a specific choice of Hamiltonian. The local density-density interaction and the single band of the Hubbard Hamiltonian bring many simplifications, but none of them are necessary for our imaginary-time integral solution, or the chemical potential tuning scheme. 

Consider the general Hamiltonian
\begin{equation}
 H = \sum_\alpha (\varepsilon_\alpha - \mu) + \sum_{\alpha_1\alpha_2\alpha_3\alpha_4} U_{\alpha_1\alpha_2\alpha_3\alpha_4} c^\dagger_{\alpha_1} c_{\alpha_2}c^\dagger_{\alpha_3} c_{\alpha_4}
\end{equation}
The $\alpha$ are the eigenstates of the non-interacting Hamiltonian, e.g. a combined momentum, band and spin index. The self-energy can be now expressed as a series
\begin{eqnarray}
 \Sigma^{(\mathrm{HF})}_{\alpha,\alpha'}(\tau) &=& 
 \sum_N \sum_{\Upsilon_N} 
  \prod_{j=1}^{2N-1} 
 \sum_{l_j=1}^{\infty} \sum_{\alpha_{j,1}...\alpha_{j,l_j}} \prod_{n=1}^{l_j-1} \sum_{\mathbf{V}_{j,n}}  
 \\ \nonumber 
 && \times [\mathbf{V}_{j,n}]_{\alpha_{j,n}\alpha_{j,n+1}} \prod_{i=1}^N  U_{\alpha_{j_1}(i) \alpha_{j_2}(i) \alpha_{j_3}(i) \alpha_{j_4}(i) }\\ \nonumber 
 && \times
 \prod_{m=1}^{N-1+\sum_j (l_j-1)} \int_0^\beta \mathrm{d}\tau_m \;
 G_0(\bar{\varepsilon}_{\alpha_{j,n}},\tilde{\tau}_{j,n}-\tilde{\tau}'_{j,n})
\end{eqnarray}
Similarly as before, $\Upsilon_N$ enumerates topologies without any instantaneous insertions (Hartree or chemical potential) at perturbation order $N$ (the number of interaction vertices).
The fermionic lines in the $\Upsilon_N$ topology are enumerated with $j$. On each fermionic line we make any number $l_j-1$ of instantaneous insertions with amplitudes $\mathbf{V}_{j,n}$ (interaction amplitudes in Hartree insertions are included in $\mathbf{V}$; $n$ enumerates the insertions at the fermionic line $j$). In general, Hartree insertions may contain off-diagonal terms in the $\alpha$-basis, and are therefore a matrix in the $\alpha$-space. However, it is necessary that chemical-potential shifts are diagonal in this basis, as we want to have the bare propagator diagonal in this basis, as well. Otherwise the form of $G_0$ from Eq.\ref{eq:g0_basic} would no longer hold.
Nevertheless, one may still have a separate chemical potential shift for each state, $\delta\mu_\alpha$.
After making insertions, the number of fermionic lines increases to $\sum_j l_j$. The fermionic lines are now enumerated with $j,n$, and the corresponding states are $\alpha_{j,n}$.
The index $i$ enumerates the interaction vertices outside of any Hartree insertions. We denote $\alpha_{j_{1..4}}(i)$ the single-particle states at 4 terminals of each interaction vertex. The interaction vertices at incoming ($i=1$) and outgoing ($i=N$) terminals of the self-energy diagram are $\alpha_{j_{1}}(i=N)=\alpha$, $\alpha_{j_{2}}(i=1)=\alpha'$. With $m$ we enumerate all times to be integrated over. With each interaction vertex $i>1$ we associate one time, and there is a time associated to each instantaneous insertion of which there are $\sum_j(l_j-1)$. We assume that the incoming time corresponding to the vertex $i=1$ is 0. The times on the terminals of each bare propagator $j,n$ are $\tilde{\tau}_{j,n}$ and $\tilde{\tau}'_{j,n}$ and they take on values from the set $\{\tau_m\}_{m=0..N-1+\sum_j (l_j-1)}$, with the external incoming time fixed, $\tau_0\equiv 0$. $\tilde{\tau}_{j,n}$,$\tilde{\tau}'_{j,n}$ and $\alpha_{j_{1..4}}(i)$ are implicit functions of topology $\Upsilon_N$. Finally, $\bar{\varepsilon}_{\alpha_{j,n}}\equiv\varepsilon_{\alpha_{j,n}}-\mu+\delta\mu_{\alpha_{j,n}}$.
We can now focus only on the time-integral part and proceed completely analogously to Eq.\ref{eq:D_starting_point}-Eq.\ref{eq:final_expression}.

It is worth noting that with general interactions, pulling the coupling constant in front of the diagram contribution is impossible, as the frequency dependence of the contribution of each diagram will depend on the precise form of $U_{\alpha_1\alpha_2\alpha_3\alpha_4}$. In the most general case, one must set specific values for $U_{\alpha_1\alpha_2\alpha_3\alpha_4}$ and $\delta\mu_\alpha$ before performing the Monte Carlo summation. One can then choose the variables that will be sampled stochastically, and the ones that will be fully summed over. In the end, the contributions can be easily grouped by total number of independent times ($K$), including those in Hartree insertions. The integration of times in Hartree insertions can always be performed beforehand. Therefore, in the fully general case, the number of integrations to be performed at the time of Monte Carlo sampling is $N-1+\sum_j(l_j-1)$. In the case of purely density-density interactions (as is the case in the Hubbard model) or spin-spin in absence of external magnetic fields, this simplifies further, because instantaneous insertions lead to expressions of type $\frac{1}{(i\omega-\varepsilon)^l}$ for which we can work out the temporal Fourier transform analytically (Eq.~\ref{eq:g0_tau}), and the remaining number of integrations to perform is $N-1$ (as we do in Eq.~\ref{eq:final_expression}). In the general case, when Hartree insertions are not diagonal in the $\alpha$-basis, one has expressions of the type $\frac{1}{i\omega-\varepsilon_1}\frac{1}{i\omega-\varepsilon_2}...\frac{1}{i\omega-\varepsilon_l}$. In principle one could prepare the analytical Fourier transforms for a general function of this form, but it might be increasingly involved at large $l$, so we assume one would do these integrations at the level of the Monte Carlo, when $\varepsilon_{1..l}$ are already specified.

We finally emphasize that even more general constructions are possible, even in bases other than the non-interacting eigenbasis. In such cases, the $G_0$'s are non-diagonal and may have a continuous real-frequency dependence, instead of being a single pole. We leave such considerations for future work.

\begin{acknowledgments}
We thank Fedor \v Simkovic for useful discussions and for sharing his diagram topology data.
Computations were performed on the PARADOX supercomputing facility (Scientific
Computing Laboratory, Center for the Study of Complex Systems, Institute of
Physics Belgrade) and the ALPHA cluster (Coll\` ege de France).  This work was
also granted access to the HPC resources of TGCC and IDRIS under the
allocations A0090510609 and A0070510609 attributed by GENCI (Grand Equipement
National de Calcul Intensif).  J.~V. acknowledges funding provided by the
Institute of Physics Belgrade, through the grant by the Ministry of Education,
Science, and Technological Development of the Republic of Serbia, as well as by
the Science Fund of the Republic of Serbia, under the Key2SM project (PROMIS
program, Grant No. 6066160).  M.~F. acknowledges support by the Simons
Foundation.  We also acknowledge support by the European Research Council for
the European Union Seventh Framework Program (FP7/2007-2013) with ERC Grant No.
319286 (QMAC).
\end{acknowledgments}

\bibliography{refs_merged.bib}
\bibliographystyle{apsrev4-1}

\end{document}